\documentclass[%
reprint,
superscriptaddress,
 amsmath,amssymb,
 aps,
 prd,
]{revtex4-2}
\usepackage[dvipsnames]{xcolor}
\usepackage{graphicx}
\usepackage{dcolumn}
\usepackage{bm}
\usepackage{subfig}
\usepackage{tabularx}
\usepackage{lineno}
\usepackage{hyperref}

\newcommand{\apjl}{Astrophys. J. Lett.}
\newcommand{\aj}{Astron. J.}
\newcommand{\mnras}{Mon. Not. R. Astron. Soc.}
\newcommand{\aap}{Astron. Astrophys. }
\newcommand{\araa}{Annu. Rev. Astron. Astrophys.}
\newcommand{\sovast}{Sov. Astron.}
\newcommand{\apss}{Astrophys. Space Sci.}
\newcommand{\ga}{{\gtrsim}}

\hypersetup{
colorlinks=true, 
linkcolor=Blue, 
citecolor=Blue, 
filecolor=black, 
urlcolor=Blue 
}
\makeatletter

\begin{document}

\preprint{APS/123-QED}

\title{Detecting Intermediate-mass Black Holes Using Miniature 
Pulsar Timing Arrays in Globular Clusters}

\author{Xian Chen}
\email{xian.chen@pku.edu.cn}
\affiliation{Department of Astronomy, School of Physics, Peking University, 100871 Beijing, China}
\affiliation{Kavli Institute for Astronomy and Astrophysics at Peking University, 100871 Beijing, China}

\author{Ver\'{o}nica V\'{a}zquez-Aceves}
\affiliation{Kavli Institute for Astronomy and Astrophysics at Peking University, 100871 Beijing, China}

\author{Siyuan Chen}
\affiliation{Shanghai Astronomical Observatory, Chinese Academy of Sciences, 80 Nandan Road, Shanghai 200030, China}
\affiliation{State Key Laboratory of Radio Astronomy and Technology, A20 Datun Road, Chaoyang District, Beijing, 100101, P. R. China}

\author{Kejia Lee}
\affiliation{Department of Astronomy, School of Physics, Peking University, 100871 Beijing, China}
\affiliation{National Astronomical Observatories, Chinese Academy of Sciences, A20 Datun Road, Chaoyang District, Beijing, 100101, P. R. China}
\affiliation{Yunnan Astronomical Observatories, Chinese Academy of Sciences, Kunming 650216, Yunnan, P. R. China}
\affiliation{Beijing Laser Acceleration Innovation Center, Huairou, Beijing, 101400, P. R.China}

\author{Yanjun Guo}
\affiliation{National Astronomical Observatories, Chinese Academy of Sciences, A20 Datun Road, Chaoyang District, Beijing, 100101, P. R. China}
\affiliation{State Key Laboratory of Radio Astronomy and Technology, A20 Datun Road, Chaoyang District, Beijing, 100101, P. R. China}

\author{Kuo Liu}
\affiliation{Shanghai Astronomical Observatory, Chinese Academy of Sciences, 80 Nandan Road, Shanghai 200030, China}
\affiliation{State Key Laboratory of Radio Astronomy and Technology, A20 Datun Road, Chaoyang District, Beijing, 100101, P. R. China}

\date{\today}

\begin{abstract}
Theoretical models predict that intermediate-mass black holes (IMBHs) exist
in globular clusters (GCs), but observational evidence remains elusive.
Millisecond pulsars (MSPs), which are abundant in GCs and have served as
precise probes for gravitational waves (GWs), offer a unique opportunity to
detect potential IMBH binaries in GCs. Here, we consider the possibility of using 
multiple MSPs in a GC to form a miniature pulsar timing array (PTA), 
so as to take advantage of their correlated timing residuals to  
	search for potential IMBH
binaries in the same cluster. Our semi-analytical
calculations reveal that nearby IMBH binaries around MSPs in GCs could induce
microsecond-level timing residuals. In GCs like $\omega$ Centauri and M15, 
favorable configurations are found which could lead to the detection of 
binaries with mass ratios $q\ga0.1$ and orbital periods of a few days. 
We estimate that future
higher-precision timing programs could achieve $100$-nanosecond sensitivity,
substantially expanding the searchable parameter space and establishing
mini-PTAs as powerful detectors of IMBHs.
\end{abstract}

\maketitle


\section{Introduction} \label{sec:intro}

Millisecond pulsars (MSPs) are highly stable rotators and can serve as precise
probes for detecting spacetime distortions caused by gravitational waves (GWs)
\cite{Sazhin_1978,Detweiler_1979}. By monitoring an array of MSPs in the
Galactic field and analyzing correlations in their timing residuals,
astronomers can detect stochastic GW backgrounds \cite{1983ApJ...265L..39H}.
This method, known as the pulsar timing array (PTA) technique, recently led to
the discovery of a possible signal of nanohertz GW background
\cite{2023ApJ...951L...9A,2023A&A...678A..50E,2023ApJ...951L...6R,2023RAA....23g5024X,2025MNRAS.536.1489M}.

Globular clusters (GCs) are one of the primary locations where MSPs are found.
Since the first GC pulsar detection \cite{1987Natur.328..399L}, over $330$
pulsars have been observed in these dense stellar systems \footnote{Pulsars in
GCs: \url{https://www3.mpifr-bonn.mpg.de/staff/pfreire/GCpsr.html}}.
Remarkably, more than $95\%$ of these GC pulsars have spin periods shorter than
$30$ milliseconds (ms), a significantly higher fraction than their counterparts
in the Galactic field.  Many are isolated, lacking binary companions (e.g.,
\cite{2023RAA....23e5012Y,yin24}). Recent GC pulsar discoveries have been made
by major radio telescopes, including FAST
\cite{2024ApJ...974L..23W,2021ApJ...915L..28P,2024ApJ...974L...2G,2025arXiv250607970L},
MeerKAT \cite{2021MNRAS.504.1407R}, GMRT \cite{2025arXiv250209154D}, and Parkes
\cite{dai20,dai23}. Future observations with the Square Kilometre Array (SKA)
are projected to detect $100–300$ pulsars in a comprehensive survey of
approximately $150$ Galactic GCs during its first phase, and the number of
pulsars may grow to $3700$ in the second phase of SKA observation
\cite{2015aska.confE..47H}.

GCs may also harbor a class of elusive objects known as intermediate-mass black
holes (IMBHs), whose masses range from $10^2$ to $10^5\,M_\odot$. While the
existence of IMBHs in GCs has long been predicted by theoretical models
\cite{2023arXiv231112118A}, robust observational evidence remains lacking (see
reviews in \cite{2017IJMPD..2630021M,2020ARA&A..58..257G}).  Nevertheless, both
numerical N-body simulations (e.g.,
\cite{2004ApJ...613.1143B,2025arXiv250322109L}) and semi-analytical Monte-Carlo
models (e.g., \cite{2015MNRAS.454.3150G}) suggest that IMBHs in GCs could
produce two distinct types of GW sources. (i) An IMBH in a GC, through frequent
gravitational interactions with surrounding stars, could capture the most
massive object in the cluster, likely a stellar-mass black hole (BH) of
$\sim10M_\odot$ \cite{2006ApJ...642..427B,2014MNRAS.444...29L}, and form a
binary of intermediate mass ratio ($10^{-4}\lesssim q\lesssim10^{-2}$).  Such a
binary can persist for a significant fraction of the cluster's lifetime
\cite{2016ApJ...819...70M,2021A&A...652A..54A} before being disrupted by
stellar interactions, replaced by another stellar companion, or driven to
merger by GW radiation \cite{2013A&A...557A.135K,2016ApJ...832..192H}. 
(ii) Two IMBHs may exist in the same GC, due to either efficient formation of
very massive stars via stellar collisions \cite{2006ApJ...640L..39G} or mergers
of GCs \cite{2006ApJ...653L..53A}. In either case, the two IMBHs could quickly
form a binary and, depending on the mass ratio and orbital eccentricity,
survive for hundreds of millions of years before merging due to GW radiation
\cite{2019arXiv190605864A,2020ApJ...899..149R,2025MNRAS.tmp..434S}.  In the
following, we will refer to the aforementioned two types of GW sources as
the ``IMBH binaries''.

The potential coexistence of MSPs and IMBH binaries within GCs has
motivated the calculation of MSP timing residuals induced by nearby GW sources.
The initial calculations focused on binary neutron stars and stellar-origin BHs
as GW sources \cite{1993Ap&SS.208...93S,1994PhRvD..50.3795F} and estimated that
the timing residual could be as large $100$ nano-seconds (ns) \cite{lommen05}.
These earlier studies, as well as the later more robust calculations
\cite{1998PhRvD..58d4003D,1999PhRvD..59h4023K,2013MNRAS.430..305H}, also
clarified a crucial difference between the current scenario and the
conventional PTA technique. In the traditional PTA setup, pulsars are far away
from the GW sources so that pulsar emissions propagate at a constant angle
relative to the GW wave vector.  However, in the context of GCs, this
assumption breaks down due to the proximity of the MSPs to the potential GW
sources.  More recently, Jenet et al. \cite{jenet05} developed a simplified
framework to calculate this ``near-field'' effect, demonstrating that the continuous GWs from IMBH binaries
can induce timing residuals with amplitudes as large as $500$ ns, well within
the detection capabilities of modern radio telescopes. Even if IMBH mergers
produce short GW bursts, their imprint on timing residuals remains detectable
when MSPs and IMBH binaries coexist in the same GCs \cite{2017PhRvD..96l3016M}.

While previous studies have primarily focused on the timing residuals of
individual MSPs, it is important to note that variations in a single pulsar's
timing residual alone cannot conclusively confirm the presence of a GW source
in the GC. Such variations could alternatively arise from the pulsar being part
of a binary system (e.g., \cite{2017MNRAS.468.2114P}).  Traditionally,
definitive evidence requires observing the characteristic angular correlation
between timing residuals of different MSPs (the Hellings-Downs curve
\citep{1983ApJ...265L..39H}).  However, this standard approach faces challenges
when applied to GC MSPs, as current observations only provide the
two-dimensional projected sky positions of MSPs, lacking the full directional
information needed for conventional correlation analysis.  Nevertheless, in
principle, the timing residuals of different MSPs within the same GC should
exhibit correlations due to their shared proximity to a potential GW source.
The key uncertainty is whether the correlations are prominent enough for detection.  

To address this question, we first derive simplified semi-analytical
expressions for near-field timing residuals based on Ref.~\cite{jenet05},
explicitly characterizing the dependence on the projected distance between the
MSP and the IMBH binary.  Next, we evaluate the sensitivity of a compact array
of MSPs in a GC, functioning as a miniature PTA, for detecting potential 
IMBH binaries within the same GC.  These theoretical developments are
timely given the recent new evidence for IMBHs in $\omega$ Centauri
\cite{Haberle_2024} and M15 \cite{2025NSRev..12..347H}, coupled with ongoing
discoveries and monitoring of new MSPs in these clusters
\cite{dai20,dai23,2024ApJ...974L..23W}.

\section{Theory}

\subsection{Timing residual}

\begin{figure}
    \centering
    \includegraphics[trim=2 1 1 1, clip,width=\linewidth]{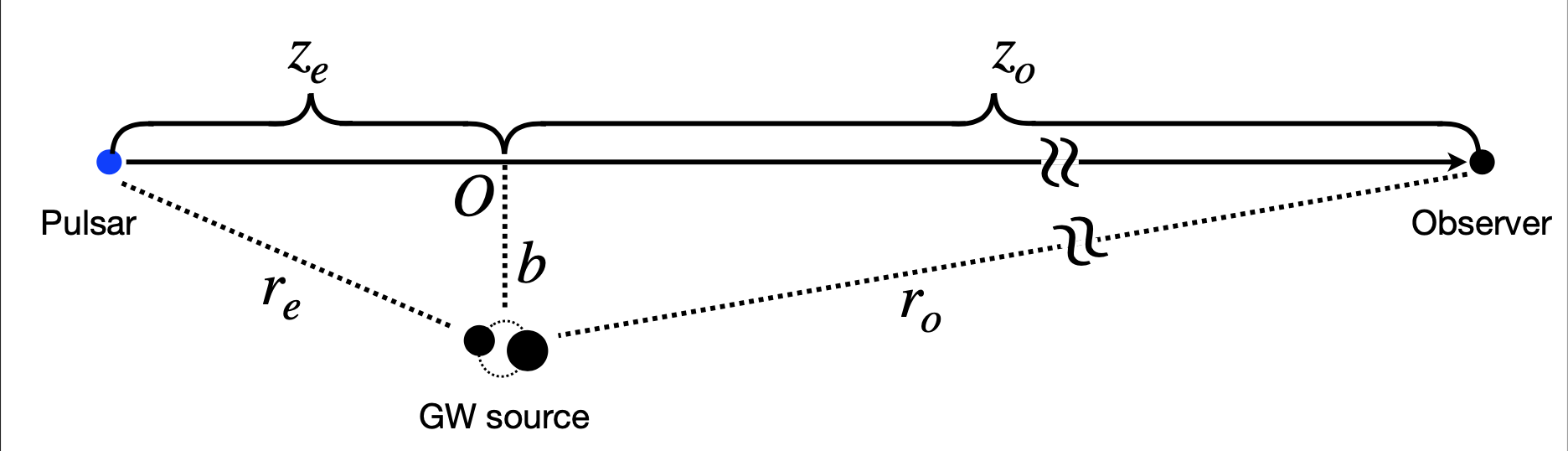}
    \caption{System configuration and definition of the parameters (also see Ref.~\cite{jenet05}).}
    \label{fig:illus}
\end{figure}

Figure~\ref{fig:illus} illustrates the system of our interest (also see Fig.~1
in Ref.~\cite{jenet05}). Both the GW source and the pulsar reside in a GC which
is far from the observer ($r_e\ll r_o$).  The GW source is a binary composed of
an IMBH of mass $m_1$ and a smaller BH of mass $m_2$.  The impact parameter $b$
quantifies the minimum distance between the GW source and the observer's line
of sight to the pulsar. We are particularly interested in the case where $b\ll
r_e$.  We define the $z$ axis to be the direction from the pulsar to the
observer (solid line with arrow).  In this direction, the pulsar's emission
intersects with the GWs from the binary source, with the angle between the
electromagnetic and GW vectors varying at each intersection point. This angular
variation distinguishes our scenario from the conventional PTA setup.

GW perturbs the spacetime at the pulsar, the observer, and in between.
As a result, the time at which a pulse arrives at the observer differs from that in flat spacetime
by an amount of
\begin{equation}
        R(t)=\int_{0}^{t}\frac{\nu_0-\nu(t')}{\nu_0}dt'\label{eq:R}
\end{equation}
\citep{Detweiler_1979,Anholm_2009}, where $\nu_0=-K^e_\mu V^{e\mu}$ is the
intrinsic spin frequency in the rest frame of the pulsar, and $\nu=-K^o_\mu
V^{o\mu}$ is the apparent spin frequency measured by the observer.  Here
$V^{\mu}$ denotes time-like four velocity, $K_\mu$ is the dual to photon
four vector, and the superscript $o$ denotes a
quantity defined at the observer while $e$ at the emitter (pulsar). The resulting
$R(t)$ is also known as the ``timing residual''.

We notice that if the pulsar or the observer is moving at a constant velocity
relative to the GW source, the equations derived below remain valid except for
a modification of the frequencies by a constant Doppler factor. Since the
velocity is typically negligible relative to the speed of light, we will ignore
it in our calculation.   If the pulsar is accelerating due to the pulling by
other stars, the sum of these accelerations may cause the apparent period of
the pulsar to vary. However, the magnitude of this acceleration is normally
smaller than $10^{-7}~{\rm m~s^{-2}}$ \cite{2017ApJ...845..148P} and the
timescale for the acceleration to change is typically $10^2-10^3$ years
\citep{1987MNRAS.225P..51B,1992RSPTA.341...39P}. Therefore, during an
observational period of, say, ten years, the second time derivate of the pulsar
period is too small to detect \citep{1987MNRAS.225P..51B} unless rare close
encounters happen \cite{1992RSPTA.341...39P}. As a result, the acceleration
induces almost a constant period derivative, which could be absorbed into the
parameters of a timing model and elude detection \citep{1987MNRAS.225P..51B}.
For this reason as well as for simplicity, we could neglect acceleration in
the current work.
 
To facilitate the later calculations, we use bars to denote the physical
quantities when there is no GW, so that $-\bar{K}_\mu
\bar{V}^{e\mu}=-\bar{K}_\mu \bar{V}^{o\mu}=\nu_0$.  When GW is introduced, we
can expand the four vectors to linear order in perturbation,
$V^{o\mu}=\bar{V}^{o\mu}+\delta V^{o\mu}$ and $K^o_\mu=\bar{K}_\mu+\delta
K^o_\mu$, and then find that  
\begin{equation} \frac{\nu_0-\nu(t)}{\nu_0} =\frac{\bar{K}_\mu\delta
V^{o\mu}+\bar{V}^{o\mu}\delta K^o_\mu}{\bar{K}^t c} \simeq\frac{\delta
K^o_t}{\bar{K}^t},\label{eq:dnu} \end{equation}
where $\bar{K}^t=\nu_0/c$.  In the last equation, we neglect the term
$\bar{K}_\mu\delta V^{o\mu}$, which is proportional to the GW amplitude at the
observer and hence proportional to $1/r_o$. While the remaining term
$\bar{V}^{o\mu}\delta K^o_\mu$ is proportional to $1/\min(r_e,b)$, therefore
bigger than the previous one by a factor of $10^4-10^5$ for typical GC systems.

To calculate $\delta K^o_t$ in Equation~(\ref{eq:dnu}), 
we assume that the GW strain tensor $h_{\mu\nu}$ is small relative to the
Minkowski metric $\eta_{\mu\nu}={\rm diag}(-1,\,1,\,1,\,1)$ and 
subtract the two null geodesic equations with and without GW. 
Then to linear order of $h_{\mu\nu}$ we derive
\begin{equation}
\frac{d\delta K_t}{d\lambda}=\frac{1}{2}h_{\mu\nu,t}\bar{K}^\mu\bar{K}^\nu,
\end{equation}
where $h_{\mu\nu,t}$ denotes the partial derivative $(\partial h_{\mu\nu}/\partial t)/c$,
and
$\lambda$ is an affine parameter for which we choose $\lambda=z/\bar{K}^t$.
Integrating the above equation from the pulsar to the observer, and noticing that
$\bar{K}^\mu=(\nu_0/c,0,0,\nu_0/c)$, we find
\begin{equation}
\frac{\delta K_t^o}{\bar{K}^t}
=\frac{\delta K_t^e}{\bar{K}^t}+
	\int_{z_e}^{z_o}\frac{h_{tt,t}+h_{zz,t}+2h_{tz,t}}{2}dz.\label{eq:dK}
\end{equation}
The first term on the right-hand side is a ``pulsar term'' caused by the 
the GWs interacting with the pulsar. The second term
results from the interaction between GW and photon.

Now the task left is to express $\delta K_t^e$ in terms of $h_{\mu\nu}$. We
start from the fact that in the rest frame of the pulsar, the spin frequency
remains $\nu_0$. Mathematically, this means $-(\bar{V}^{e\mu}+\delta
V^{e\mu})(\bar{K}^e_{\mu}+\delta K^e_{\mu})=\nu_0$.  Assuming that the pulsar is
not moving, i.e., $\bar{V}^{e\mu}=(c,0,0,0)$, to linear order we can write
\begin{equation}
\delta K^e_t=-\frac{\nu_0}{c^2}\left(\delta V^{e}_t+\delta V^{e}_z\right).\label{eq:dKe}
\end{equation}
Applying the geodesic equation for the pulsar, 
\begin{equation} 
        \frac{d\delta V_\alpha}{d\tau}=\frac{1}{2}h_{\mu\nu,\alpha}\bar{V}^\mu\bar{V}^\nu
        =\frac{c^2}{2}h_{tt,\alpha},\label{eq:dV}
\end{equation}
we finally find 
\begin{equation}
	\frac{\delta K^e_t}{\bar{K}^t}=-c\int_0^{t}\frac{h_{tt,t}+h_{tt,z}}{2}dt'.\label{eq:dKet}
\end{equation}
Notice that Equation~(\ref{eq:dKet}) is integrated over time, while the
integration in Equation~(\ref{eq:dK}) is performed with respect to space.  

\subsection{GW strain}

Now we evaluate the GW strain $h_{\mu\nu}$ 
at the position $\mathbf{r}$ and the local time $t$. 
We follow Ref.~\cite{jenet05} and use the formula
\begin{eqnarray}
h_{\mu\nu}(t,\mathbf{r})=\frac{4G}{c^4}
	\int\frac{\bar{T}_{\mu\nu}(t-|\mathbf{r}-\mathbf{r}'|/c,\mathbf{r}')}
	{|\mathbf{r}-\mathbf{r}'|}d^3\mathbf{r}',
\end{eqnarray}
where
$\bar{T}_{\mu\nu}$ is the trace-reversed stress energy tensor for the GW source, 
$\mathbf{r}'$ refers to the source position,
and
notice that the result is not in transverse-traceless gauge.
Because $r'=|\mathbf{r}'|$ is comparable to the size of the GW source, and in our scenario
much smaller than $r=|\mathbf{r}|$, we can expand $|\mathbf{r}-\mathbf{r}'|$
to the order of $O(r')$ and rewrite the strain as
\begin{eqnarray}
h_{\mu\nu}(t,\mathbf{r})
\simeq\frac{4G}{rc^4}\int\left(
	\bar{T}_{\mu\nu}+\bar{T}_{\mu\nu}\frac{\mathbf{n}\cdot\mathbf{r}'}{r}
	+\frac{\mathbf{n}\cdot\mathbf{r}'}{c}\frac{\partial \bar{T}_{\mu\nu}}{\partial t}
	\right)d^3\mathbf{r}',\label{eq:happrox}
\end{eqnarray}
where $\bar{T}_{\mu\nu}(t-r/c,\mathbf{r}')$ and $\mathbf{n}=\mathbf{r}/r$.
The second term in the integrand is smaller than the first one by 
a factor of $r'/r$, and hence we neglect it in the later calculation.
The third term is also negligible in the low-velocity limit, which 
applies to our GW source.

Therefore, we can keep only the first term in the integrand of
Equation~(\ref{eq:happrox}).  Then applying the standard perturbative technique in
the weak-field, low-velocity limit \citep{maggiore07},  we derive 
\begin{eqnarray}
	&&h_{tt}=\frac{2G}{rc^4}\left(Mc^2
	+\frac{\ddot{Q}_{xx}+\ddot{Q}_{yy}+\ddot{Q}_{zz}}{2}\right),\label{eq:htt}\\
	&&h_{zz}=\frac{2G}{rc^4}\left(Mc^2
	-\frac{\ddot{Q}_{xx}+\ddot{Q}_{yy}-\ddot{Q}_{zz}}{2}\right),\label{eq:hzz}\\
	&&h_{tz}=-\frac{4GP^z}{rc^4}=0,\label{eq:htz}
\end{eqnarray}
where $M=\int (T^{00}/c^2)d^3\mathbf{r}'=m_1+m_2$ is the total mass of the binary GW source, $Q_{ij}=Q^{ij}=\int T^{00}x'^i x'^jd^3\mathbf{r}'$ is the mass quadrupole moment of the source,
and $P^z=\int T^{0z}d^3\mathbf{r}'$ is the linear momentum in the $z$ direction. 
For the convenience of later calculation, we define $Q\equiv Q_{xx}+Q_{yy}+Q_{zz}$.

When using Equations~(\ref{eq:htt})-(\ref{eq:htz}), one has to treat $Q_{ij}$ 
as a function of the retarded time $t_*(t,z) =t-(z_o-z)/c-\sqrt{z^2+b^2}/c$, so that
$t$ corresponds to the time when the pulse arrives at the observer. 
Consequently, both $Q_{ij}$ and $r=\sqrt{z^2+b^2}$ are functions of $z$, so we can 
derive the following useful relation
\begin{eqnarray}
        h_{tt,z}=-\frac{z\,h_{tt}}{z^2+b^2}+h_{tt,t}\left(1-\frac{z}{r}\right).
\end{eqnarray}
Using this relation, as well as Equations~(\ref{eq:dnu}), (\ref{eq:dK}), and (\ref{eq:dKet}),
we find that
\begin{eqnarray}
\frac{\nu_0-\nu(t)}{\nu_0}&
=&\int_0^t \frac{z_e\,h_{tt}}{2r_e^2}\,c dt'
-\int_0^t\frac{h_{tt,t}}{2}\left(2-\frac{z_e}{r_e}\right)\,c dt'\nonumber\\
	&+&\int_{z_e}^{z_o}\frac{h_{tt,t}+h_{zz,t}}{2}dz.\label{eq:master}
\end{eqnarray}

\subsection{Semi-analytical expressions}\label{sec:domi}

Further calculation of $R(t)$ requires some specification of the system.
Without loss of generality, we assume that the GW source, i.e., the IMBH binary, 
is circular with an orbital angular frequency of $\omega_b$. The
resulting GW radiation is monochromatic with a frequency of $\omega_b/\pi$.
Since GWs from eccentric binaries can be decomposed into a series of harmonics
\cite{Peters_1963}, our analysis of single harmonic components provides a
foundation for future studies of eccentric binary GW sources.

Motivated by observations, we consider an IMBH of a mass of $m_1=(10^3-10^4)M_\odot$,
and we assume that it resides at the center of a GC. The companion is a BH of equal or smaller
mass, so that
the mass ratio of the binary is $q=m_2/m_1\le1$. The semimajor axis of the binary orbit
is typically $a=(10^2-10^7)r_g$, where $r_g=Gm_1/c^2$ is the gravitational radius of the IMBH. 
Therefore, the orbital period is
\begin{eqnarray}
	P=2\pi\left(\frac{a^3}{GM}\right)^{1/2}\simeq
\frac{113\,{\rm days}}{(1+q)^{1/2}}
	\left(\frac{m_1}{10^4M_\odot}\right)\left(\frac{a}{10^5r_g}\right)^{3/2},
\end{eqnarray}
and due to GW radiation \cite{Peters_1963} the binary has a lifetime of
\begin{eqnarray}
T_{\rm gw}&=&\frac{a}{4|\dot{a}|}=\frac{5a^4}{256cr_g^3q(1+q)}\\
	&\simeq&\frac{3.05\times10^9\,{\rm yrs}}{q(1+q)}\left(\frac{m_1}{10^4M_\odot}\right)
	\left(\frac{a}{10^5r_g}\right)^4.\label{eq:Tgw}
\end{eqnarray}

Given such a circular binary system with a reduced mass of $\mu=m_1m_2/(m_1+m_2)$ and semimajor axis of $a$, 
the mass quadrupole moment is of the order of
$\sim\mu a^2\cos(2\omega_b t+\phi)$, where the exact coefficient and the phase
($\phi$) depend on the orientation of the binary.
Since a summation of sinusoidal functions of the same frequency remains a
sinusoidal function except for 
a different coefficient and an additional phase factor, we can regard the mass quadrupole moments
and their linear combinations as functions of 
$\cos(2\omega_b t+\phi)$.  Their time derivatives can be derived
accordingly, e.g., $\ddot{Q}\sim -4\omega_b^2\mu
a^2\cos(2\omega_b t+\phi)$.

Given the above specification, we can integrate the first term in Equation~(\ref{eq:master}).
In the integrand, the term containing $M/r_e$ does not vary with time; therefore,
the integral can be absorbed into a constant Doppler shift in the timing model.
Integrating the rest time-dependent parts gives
\begin{eqnarray}
        \frac{z_eG}{2r_e^3c^3}\int_0^t \ddot{Q} dt'
	=A_1\left(\frac{\omega_b \mu a^2z_eG}{{r_e}^3c^3}\right)&\cos(2\omega_b t+\phi_1)&\label{eq:term1}\\
	=\frac{A_1 q}{(1+q)^{1/2}}
\sqrt{\frac{a}{r_g}}
\left(\frac{z_e}{r_e}\right)
        \left(\frac{r_g}{r_e}\right)^2&\cos(2\omega_b t+\phi_1)&,
\end{eqnarray}
where the coefficient $A_1$ is of order unity, and the phase $\phi_1$ depends on the orientation of the binary as well as the initial position of the companion BH.  If we use $M=10^4M_\odot$, $q\simeq1$, $z_e\sim
r_e$, $a\sim10^5r_g$, and $r_e\sim1$ pc as the fiducial values, we find that the magnitude of
$(\nu_0-\nu)/\nu_0$ contributed by the first term in Equation~(\ref{eq:master})
is about $10^{-17}$. To estimate the corresponding timing residual, 
we notice that $1/(2\omega_b)\sim10^6$ s in the fiducial model. 
Therefore, $R(t)$ is of the order of
$10^{-11}$ s due to the first term, which is too small for the current
radio telescopes to detect.

As for the second term in Equation~(\ref{eq:master}), the integration gives
\begin{eqnarray}
\frac{G\ddot{Q}}{2r_ec^4}\left(\frac{z_e}{r_e}-2\right)
	=&&A_2\left(\frac{2qr_g^2}{ar_e}\right)
\left(\frac{z_e}{r_e}-2\right)\nonumber\\
	&&\times\cos(2\omega_bt+\phi_2).
\end{eqnarray}
In our fiducial model, it is a factor of
\begin{equation}
\frac{2r_e r_g^{1/2}}{(1+q)^{1/2}a^{3/2}}\sim10^2
\end{equation}
times bigger than the first term in Equation~(\ref{eq:master}). The corresponding
$R(t)$ reaches a few ns, but still too small to detect.

Finally, we integrate the third term in Equation~(\ref{eq:master}).
With Equations~(\ref{eq:htt}) and (\ref{eq:hzz}), the integral reduces to
\begin{equation}
	\int_{z_e}^{z_o}\frac{G\dddot{Q}_{zz}}{rc^5}dz
= A_3\left(\frac{8G\mu a^2\omega_b^3}{c^5}\right)\left.H(t,z)\right|_{z_e}^{z_o},\label{eq:term3}
\end{equation}
where
\begin{eqnarray}
	&H&(t,z)=\int^z\frac{\sin(2\omega_b t_*+\phi_3)}{\sqrt{{z'}^2+b^2}}dz'\\
&=& \cos\left[2\omega_b (t-z_o/c+\phi_3)\right]{\rm Si}\left[\frac{\sqrt{z^2+b^2}-z}{c/(2\omega_b)}\right]\nonumber\\
	&-&\sin\left[2\omega_b (t-z_o/c+\phi_3)\right]{\rm Ci}\left[\frac{\sqrt{z^2+b^2}-z}{c/(2\omega_b)}\right],\label{eq:H}
\end{eqnarray}
and 
\begin{eqnarray}
	{\rm Si}(x)&=&\int^{x} \frac{\sin(x')}{x'}dx',\\
	{\rm Ci}(x)&=&\int^{x} \frac{\cos(x')}{x'}dx'.
\end{eqnarray}
We notice that the fraction in the parentheses of Equation~(\ref{eq:term3}),
\begin{eqnarray}
	\frac{8G\mu a^2\omega_b^3}{c^5}=8 q(1+q)^{1/2}\left(\frac{r_g}{a}\right)^{5/2},
\end{eqnarray}
is bigger than the fraction in Equation~(\ref{eq:term1}) by a factor of
\begin{equation}
8(1+q)\frac{r_g r_e^3}{a^3 z_e}\sim10^4
\end{equation}
in our fiducial model. 

Therefore, the third term on the right-hand side of
Equation~(\ref{eq:master}) can potentially dominate the first two terms.
This term originates from
the interaction between electromagnetic pulses and GWs. Since the integral depends on
$H(t,z)$ which is a sinusoidal function of $2\omega_bt$, we can integrate it
with respect to $t$ according to Equation~(\ref{eq:R}), and estimate the
magnitude of the timing residual as
\begin{eqnarray}
	&&\frac{R(t)}{\cos(2\omega_b t+\phi)}
	\sim  \left(\frac{4G\mu a^2\omega_b^2}{c^5}\right)\sqrt{(\Delta{\rm Ci})^2+(\Delta{\rm Si})^2}\\
&&\simeq2\mu {\rm s}~\sqrt{(\Delta{\rm Ci})^2+(\Delta{\rm Si})^2}
\left(\frac{m_1q}{10^4M_\odot}\right)
	\left(\frac{a}{10^5r_g}\right)^{-1},\label{eq:A}
\end{eqnarray}
where $\Delta{\rm Ci}$ ($\Delta{\rm Si}$) represents the difference between the two
cosine (sine) integrals evaluated at the pulsar and the observer positions.
These differences arise from the calculation of $H(t,z)|_{z_e}^{z_o}$ in
Equation~(\ref{eq:term3}). 

Interestingly, we find that the value of $\sqrt{(\Delta{\rm Ci})^2+(\Delta{\rm
Si})^2}$ is of the order of unity for our fiducial GC-MSP systems.  This is
because the value of the term $(\sqrt{z^2+b^2}-z)/(c/2\omega_b)$ in the
brackets of Equation~(\ref{eq:H}) is typically $\gg1$ at the emitter ($z=z_e$)
and, more importantly, $\ll1$ at the observer ($z=z_o\sim10$ kpc).  The notably small value
observed at $z=z_o$ is primarily due to the small impact parameter $b$, which
satisfies the condition $b/z_o\ll1$.  As a result, both $|\Delta{\rm Ci}|$ and
$|\Delta{\rm Si}|$ turn out to be of order unity. Another interesting and useful
result is that 
in real GC systems where $(\sqrt{z_e^2+b^2}-z_e)/(c/2\omega_b)\gg1$ is often satisfied,
neither $|\Delta{\rm Ci}|$ nor
$|\Delta{\rm Si}|$ is sensitive to the sign or magnitude of $z_e$. 
In this case,
the timing residual $R(t)$ only depends on the projected distance $b$
between the pulsar and the IMBH binary, making observational constraints more feasible.

\section{Examples}\label{sec:examp}

The coefficient of $2\mu{\rm s}$ in Equation~(\ref{eq:A}) puts the timing residual in a
potentially detectable range.  In the following, we showcase two real GC-MSP
systems to highlight the detectability of the timing residual.

\textbf{$\omega$ Cen} is the most massive GC in the Galaxy and one of the best
observationally studied due to its closeness to the earth ($5.2$ kpc).  It has
been considered a prime candidate for hosting an IMBH in its core
\cite{2020ARA&A..58..257G}. Recent discovery of several fast-moving stars
within its central $0.1$ pc further suggests that the IMBH should be
$\gtrsim10^4M_\odot$. However, the distribution and acceleration of the MSPs in
$\omega$ Cen seems more consistent with an extended central mass, and the
corresponding models have put an upper limit of about $6\times10^3M_\odot$ on
the IMBH mass \cite{Banares_2024}.

\begin{figure}
    \centering
    \includegraphics[width=\linewidth]{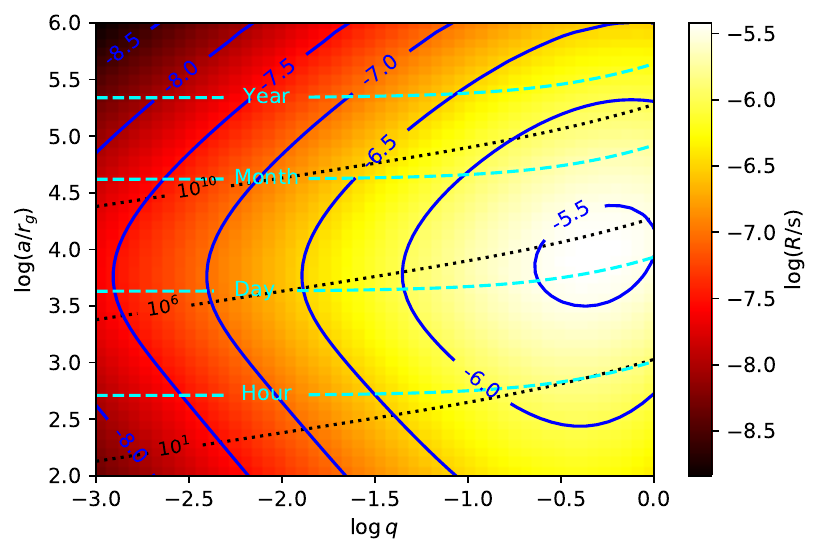}
	\caption{Timing residual (color map and blue contours) for a MSP in 
$\omega$ Cen, where we have assumed an impact parameter of $b=1$ pc for the MSP
and a total mass of $m_1+m_2=10^4M_\odot$
for the binary GW source. The four cyan dashed curves correspond to orbital periods ranging from $1$ year to $1$ hour, and the black dotted curves correspond to three GW radiation timescales, $T_{\rm gw}=10^{10}$, $10^6$, and $1$ yr, for the binary.}
    \label{fig:Rb10}
\end{figure}

Given the current theoretical and observational uncertainties, 
we assume $m_1+m_2=10^4M_\odot$, as well as $b=1$ pc and $z_e=-1$ pc to account for a typical MSP in
$\omega$ Cen \cite{dai20}.  The resulting timing residual as a function of the
binary semimajor axis $a$ and mass ratio $q$ is shown in Figure~\ref{fig:Rb10}.
We find that $R(t)$ reaches a maximum value of about $3\mu{\rm s}$ when the two
BHs are relatively equal in mass ($q\gtrsim0.3$) and on an orbit of a period of
about a few days. This maximum is comparable to the coefficient $2\mu{\rm s}$,
indicating that $H(t,z)|_{z_e}^{z_o}\sim1$.

Current observations of the MSPs in $\omega$ Cen have achieved a timing
precision of $\sim10$ $\mu{\rm s}$ with an integration time of $\sim1$ hour \cite{dai23}.
If the
same pulsars are observed repeatedly, accumulating  $N\sim10^2$ timing residual
measurements, it becomes feasible to detect a residual as small as
$\sim1\mu{\rm s}$ (see Section~\ref{sec:det}).  According to
Figure~\ref{fig:Rb10}, those GW sources resulting in $R>1\mu{\rm s}$ mostly
have $q>0.1$ and distribute in a broad semimajor-axis range. The widest
binaries among them have a lifetime of $10^{10}$ yrs, increasing the
probability that such binaries may exist in $\omega$ Cen today.

\textbf{M15} is a core-collapse GC located at a distance of $10$ kpc from the earth.
Current stellar kinematic models remain inconclusive regarding the existence
and mass of an IMBH in M15, with possible mass estimates ranging from about
$4000M_\odot$ to as small as $500M_\odot$
\cite{2002AJ....124.3270G,2003ApJ...582L..21B,2025NSRev..12..347H}.  Meanwhile,
$15$ pulsars have been discovered in M15 where more than half of them have a
spin period shorter than $30\mu{\rm s}$.  In particular, a few MSPs reside
within a projected distance of $0.1$ pc from the center of the GC, 
and recent timing of them has
achieved a precision of about $20\mu{\rm s}$
\cite{2024ApJ...974L..23W}.

\begin{figure}
    \centering
    \includegraphics[width=\linewidth]{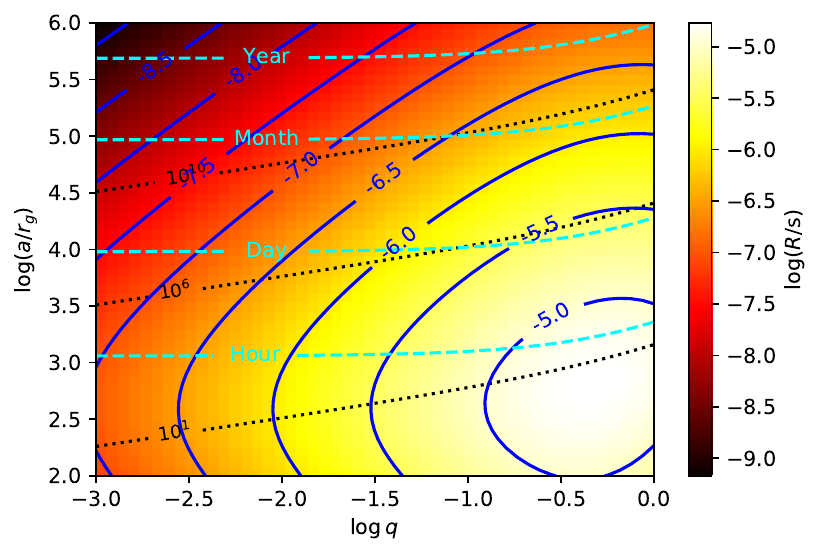}
\caption{Same as Figure~\ref{fig:Rb10} but for M15, where we have assumed $b=0.1$ pc and $m_1+m_2=3\times10^3M_\odot$.}
    \label{fig:Rb01}
\end{figure}

Therefore, we assume a total mass of $m_1+m_2=3\times10^3M_\odot$ for the IMBH
binary in M15 \cite{2002AJ....124.3270G}, and $b=0.1$ pc and $z_e=-1$ pc for
the MSP \cite{2024ApJ...974L..23W}.  The resulting timing residual as a
function of the binary semimajor axis $a$ and mass ratio $q$ is shown in
Figure~\ref{fig:Rb01}. Now the peak of $R(t)$ is higher than in the previous
case, exceeding $10\mu{\rm s}$. But the location of the peak in the $q-a$
diagram shifts to smaller $a/r_g$. Because the corresponding orbital period is
shorter than an hour, detecting such a GW source requires high observational
cadence, which is challenging with the current telescopes. However, a fraction
of the parameter space above the curve of $P=1$ day still allows $R(1)>1\mu{\rm
s}$.  Here, IMBH binaries have a lifetime between $10^6$ and $10^9$ years (see
black dotted lines).  Whether or not such binaries exist can be tested by
timing the MSPs in M15.

\section{Detectability}\label{sec:det}

Now we discuss how to find a signal like that described in Equation~(\ref{eq:A}) from noisy timing data. Suppose the noise 
$n(t)$ follows a Gaussian distribution with zero mean 
$\left<n\right>=0$ and has a variance of
${\rm Var}(n)=\left<n^2\right>=\sigma^2$. 
In the null hypothesis, when there is no signal, 
the
data $d(t_i)$ will contain pure noise,
where $t_i$ denotes the time when the $i$th timing residual is measured.
If one uses a
template $R(t)=\cos(\tilde{\omega} t_i+\tilde{\phi})$ to do a matched-filtering
analysis, i.e., calculating the overlap
\begin{equation}
	T_n=\sum_{i=1}^{N} d(t_i) R(t_i),\label{eq:T}
\end{equation}
the result will have a mean value of $\bar{T}_n=0$ with a variance of ${\rm
Var}(T_n)=N\sigma^2/2$.

In the alternative hypothesis, if there is a sinusoidal signal of amplitude $A$,
angular frequency $\omega$,
and phase $\phi$,
the data will become $d(t_i)=s(t_i)+n(t_i)$, where
$s(t_i)=A\cos(\omega t_i+\phi)$ is the signal. 
Notice that we are interested in the case $A\ll \sigma$.  
Now the result of  Equation~(\ref{eq:T}) will depend on the choices of $\tilde{\omega}$
and $\tilde{\phi}$. But one can experiment with different $\tilde{\omega}$
and $\tilde{\phi}$ to look for the maximum, which we denote as  $T_{\rm max}$. 
In the ideal situation, that the number of measurements $N$
is sufficiently large, and the time span of the data covers at least one
GW period $2\pi/\omega$, the maximum should be found at
$\tilde{\omega}=\omega$ and $\tilde{\phi}=\phi$, and  the corresponding value is
\begin{equation} T_{\rm max}=\sum_{i=1}^{N} A\cos^2(\omega
t_i+\phi)+\sum_{i=1}^{N}n_i\cos(\omega t_i+\phi). \end{equation}
The mean and the variance of $T_{\rm max}$ are, respectively,
\begin{eqnarray} \bar{T}_{\rm max}&\simeq& \frac{NA}{2},\\ {\rm
	Var}(T_{\rm max})&\simeq&\frac{NA^2}{8}+\frac{N\sigma^2}{2}\simeq\frac{N\sigma^2}{2},
\end{eqnarray}
where, again, we have assumed a long observation duration and a large number of
timing residual measurements, so that those terms containing $\cos(\omega t_i+\phi)$,
$\cos(2\omega t_i+2\phi)$, $\cos(3\omega t_i+3\phi)$, and $\cos(4\omega t_i+4\phi)$
all vanish upon ensemble averaging.

Now we propose a detection threshold based on two considerations.
(1) Signal presence condition: 
If a true signal exists, $T_{\rm max}$ should exceed
$\bar{T}_{\rm max}-1.64\sqrt{{\rm Var}(T_{\rm max})}$ with $95\%$ probability. 
(2) Statistical significance condition:  
To limit false alarms to less than $5\%$, we require 
$T_{\rm max}>\bar{T}_n+1.64\sqrt{{\rm Var}(T_n)}$.
The combined threshold is therefore
$\bar{T}_{\rm max}-1.64\sqrt{{\rm Var}(T_{\rm max})}>\bar{T}_n+1.64\sqrt{{\rm Var}(T_n)}$,
which corresponds to
\begin{equation}
A\ga4.65\sigma/\sqrt{N}.
\end{equation}
Signals with this amplitude can be reliably distinguished from background noise
with high probability.

Current observations of MSPs in GCs are approaching a timing precision
(weighted rms) of $\sigma\simeq10\mu{\rm s}$ as well as a total number of $100$ measurements for
a single MSP \cite{dai23,2024ApJ...974L..23W}. They will allow us to detect GWs
with an amplitude of $A\ga 4.7\mu{\rm s}$.  Such a sensitivity corresponds to a
contour level of about $-5.3$ in Figures~\ref{fig:Rb10} and \ref{fig:Rb01}.
Interestingly, such contours will enclose a small, but non-zero fraction of the parameter space. 
Therefore, we would be able to test the existence of IMBH binaries in GCs by timing even a single MSP.

More importantly, as we have mentioned earlier, multiple MSPs are often found
in a single GC, and the number $N_p$ could be as large as $O(10)$.  According to
our analysis of the matched-filtering method, using the timing data of all the
MSPs with similar timing precision will further improve the sensitivity by a
factor of $\sqrt{N_p}$.  It is worth mentioning that by substantially
increasing the timing precision ($\sigma$) or the number of MSPs ($N_p$), a sensitivity of $A\sim
O(10^2)$ ns may be reached in the future.  Then a significantly larger portion
of the parameter space shown in Figures~\ref{fig:Rb10} and \ref{fig:Rb01} for
potential IMBH binaries could become accessible via pulsar timing.  The
envisioned mini-PTA will then serve as an even more powerful probe for IMBHs.

\begin{acknowledgments}
This work is supported by the National Key Research and Development Program of
China (Grant No. SQ2024YFC220046) and the National Natural Science Foundation of
China (Grant No. 12473037). KJL thanks the Xplorer Prize for the support.
\end{acknowledgments}

\bibliography{ptabib}

\begin{thebibliography}{58}%
\makeatletter
\providecommand \@ifxundefined [1]{%
 \@ifx{#1\undefined}
}%
\providecommand \@ifnum [1]{%
 \ifnum #1\expandafter \@firstoftwo
 \else \expandafter \@secondoftwo
 \fi
}%
\providecommand \@ifx [1]{%
 \ifx #1\expandafter \@firstoftwo
 \else \expandafter \@secondoftwo
 \fi
}%
\providecommand \natexlab [1]{#1}%
\providecommand \enquote  [1]{``#1''}%
\providecommand \bibnamefont  [1]{#1}%
\providecommand \bibfnamefont [1]{#1}%
\providecommand \citenamefont [1]{#1}%
\providecommand \href@noop [0]{\@secondoftwo}%
\providecommand \href [0]{\begingroup \@sanitize@url \@href}%
\providecommand \@href[1]{\@@startlink{#1}\@@href}%
\providecommand \@@href[1]{\endgroup#1\@@endlink}%
\providecommand \@sanitize@url [0]{\catcode `\\12\catcode `\$12\catcode
  `\&12\catcode `\#12\catcode `\^12\catcode `\_12\catcode `\%12\relax}%
\providecommand \@@startlink[1]{}%
\providecommand \@@endlink[0]{}%
\providecommand \url  [0]{\begingroup\@sanitize@url \@url }%
\providecommand \@url [1]{\endgroup\@href {#1}{\urlprefix }}%
\providecommand \urlprefix  [0]{URL }%
\providecommand \Eprint [0]{\href }%
\providecommand \doibase [0]{https://doi.org/}%
\providecommand \selectlanguage [0]{\@gobble}%
\providecommand \bibinfo  [0]{\@secondoftwo}%
\providecommand \bibfield  [0]{\@secondoftwo}%
\providecommand \translation [1]{[#1]}%
\providecommand \BibitemOpen [0]{}%
\providecommand \bibitemStop [0]{}%
\providecommand \bibitemNoStop [0]{.\EOS\space}%
\providecommand \EOS [0]{\spacefactor3000\relax}%
\providecommand \BibitemShut  [1]{\csname bibitem#1\endcsname}%
\let\auto@bib@innerbib\@empty
\bibitem [{\citenamefont {{Sazhin}}(1978)}]{Sazhin_1978}%
  \BibitemOpen
  \bibfield  {author} {\bibinfo {author} {\bibfnamefont {M.~V.}\ \bibnamefont
  {{Sazhin}}},\ }\bibfield  {title} {\bibinfo {title} {{Opportunities for
  detecting ultralong gravitational waves}},\ }\href@noop {} {\bibfield
  {journal} {\bibinfo  {journal} {\sovast}\ }\textbf {\bibinfo {volume} {22}},\
  \bibinfo {pages} {36} (\bibinfo {year} {1978})}\BibitemShut {NoStop}%
\bibitem [{\citenamefont {{Detweiler}}(1979)}]{Detweiler_1979}%
  \BibitemOpen
  \bibfield  {author} {\bibinfo {author} {\bibfnamefont {S.}~\bibnamefont
  {{Detweiler}}},\ }\bibfield  {title} {\bibinfo {title} {{Pulsar timing
  measurements and the search for gravitational waves}},\ }\href
  {https://doi.org/10.1086/157593} {\bibfield  {journal} {\bibinfo  {journal}
  {\apj}\ }\textbf {\bibinfo {volume} {234}},\ \bibinfo {pages} {1100}
  (\bibinfo {year} {1979})}\BibitemShut {NoStop}%
\bibitem [{\citenamefont {{Hellings}}\ and\ \citenamefont
  {{Downs}}(1983)}]{1983ApJ...265L..39H}%
  \BibitemOpen
  \bibfield  {author} {\bibinfo {author} {\bibfnamefont {R.~W.}\ \bibnamefont
  {{Hellings}}}\ and\ \bibinfo {author} {\bibfnamefont {G.~S.}\ \bibnamefont
  {{Downs}}},\ }\bibfield  {title} {\bibinfo {title} {{Upper limits on the
  isotropic gravitational radiation background from pulsar timing analysis.}},\
  }\href {https://doi.org/10.1086/183954} {\bibfield  {journal} {\bibinfo
  {journal} {\apjl}\ }\textbf {\bibinfo {volume} {265}},\ \bibinfo {pages}
  {L39} (\bibinfo {year} {1983})}\BibitemShut {NoStop}%
\bibitem [{\citenamefont {{Agazie}}\ \emph {et~al.}(2023)\citenamefont
  {{Agazie}}, \citenamefont {{Alam}}, \citenamefont {{Anumarlapudi}},
  \citenamefont {{Archibald}}, \citenamefont {{Arzoumanian}}, \citenamefont
  {{Baker}}, \citenamefont {{Blecha}}, \citenamefont {{Bonidie}}, \citenamefont
  {{Brazier}}, \citenamefont {{Brook}}, \citenamefont {{Burke-Spolaor}},
  \citenamefont {{B{\'e}csy}}, \citenamefont {{Chapman}}, \citenamefont
  {{Charisi}}, \citenamefont {{Chatterjee}}, \citenamefont {{Cohen}},
  \citenamefont {{Cordes}}, \citenamefont {{Cornish}}, \citenamefont
  {{Crawford}}, \citenamefont {{Cromartie}}, \citenamefont {{Crowter}},
  \citenamefont {{Decesar}}, \citenamefont {{Demorest}}, \citenamefont
  {{Dolch}}, \citenamefont {{Drachler}}, \citenamefont {{Ferrara}},
  \citenamefont {{Fiore}}, \citenamefont {{Fonseca}}, \citenamefont
  {{Freedman}}, \citenamefont {{Garver-Daniels}}, \citenamefont {{Gentile}},
  \citenamefont {{Glaser}}, \citenamefont {{Good}}, \citenamefont
  {{G{\"u}ltekin}}, \citenamefont {{Hazboun}}, \citenamefont {{Jennings}},
  \citenamefont {{Jessup}}, \citenamefont {{Johnson}}, \citenamefont {{Jones}},
  \citenamefont {{Kaiser}}, \citenamefont {{Kaplan}}, \citenamefont {{Kelley}},
  \citenamefont {{Kerr}}, \citenamefont {{Key}}, \citenamefont {{Kuske}},
  \citenamefont {{Laal}}, \citenamefont {{Lam}}, \citenamefont {{Lamb}},
  \citenamefont {{Lazio}}, \citenamefont {{Lewandowska}}, \citenamefont
  {{Lin}}, \citenamefont {{Liu}}, \citenamefont {{Lorimer}}, \citenamefont
  {{Luo}}, \citenamefont {{Lynch}}, \citenamefont {{Ma}}, \citenamefont
  {{Madison}}, \citenamefont {{Maraccini}}, \citenamefont {{McEwen}},
  \citenamefont {{McKee}}, \citenamefont {{McLaughlin}}, \citenamefont
  {{McMann}}, \citenamefont {{Meyers}}, \citenamefont {{Mingarelli}},
  \citenamefont {{Mitridate}}, \citenamefont {{Ng}}, \citenamefont {{Nice}},
  \citenamefont {{Ocker}}, \citenamefont {{Olum}}, \citenamefont {{Panciu}},
  \citenamefont {{Pennucci}}, \citenamefont {{Perera}}, \citenamefont {{Pol}},
  \citenamefont {{Radovan}}, \citenamefont {{Ransom}}, \citenamefont {{Ray}},
  \citenamefont {{Romano}}, \citenamefont {{Salo}}, \citenamefont {{Sardesai}},
  \citenamefont {{Schmiedekamp}}, \citenamefont {{Schmiedekamp}}, \citenamefont
  {{Schmitz}}, \citenamefont {{Shapiro-Albert}}, \citenamefont {{Siemens}},
  \citenamefont {{Simon}}, \citenamefont {{Siwek}}, \citenamefont {{Stairs}},
  \citenamefont {{Stinebring}}, \citenamefont {{Stovall}}, \citenamefont
  {{Susobhanan}}, \citenamefont {{Swiggum}}, \citenamefont {{Taylor}},
  \citenamefont {{Turner}}, \citenamefont {{Unal}}, \citenamefont
  {{Vallisneri}}, \citenamefont {{Vigeland}}, \citenamefont {{Wahl}},
  \citenamefont {{Wang}}, \citenamefont {{Witt}}, \citenamefont {{Young}},\
  and\ \citenamefont {{Nanograv Collaboration}}}]{2023ApJ...951L...9A}%
  \BibitemOpen
  \bibfield  {author} {\bibinfo {author} {\bibfnamefont {G.}~\bibnamefont
  {{Agazie}}}, \bibinfo {author} {\bibfnamefont {M.~F.}\ \bibnamefont
  {{Alam}}}, \bibinfo {author} {\bibfnamefont {A.}~\bibnamefont
  {{Anumarlapudi}}}, \bibinfo {author} {\bibfnamefont {A.~M.}\ \bibnamefont
  {{Archibald}}}, \bibinfo {author} {\bibfnamefont {Z.}~\bibnamefont
  {{Arzoumanian}}}, \bibinfo {author} {\bibfnamefont {P.~T.}\ \bibnamefont
  {{Baker}}}, \bibinfo {author} {\bibfnamefont {L.}~\bibnamefont {{Blecha}}},
  \bibinfo {author} {\bibfnamefont {V.}~\bibnamefont {{Bonidie}}}, \bibinfo
  {author} {\bibfnamefont {A.}~\bibnamefont {{Brazier}}}, \bibinfo {author}
  {\bibfnamefont {P.~R.}\ \bibnamefont {{Brook}}}, \bibinfo {author}
  {\bibfnamefont {S.}~\bibnamefont {{Burke-Spolaor}}}, \bibinfo {author}
  {\bibfnamefont {B.}~\bibnamefont {{B{\'e}csy}}}, \bibinfo {author}
  {\bibfnamefont {C.}~\bibnamefont {{Chapman}}}, \bibinfo {author}
  {\bibfnamefont {M.}~\bibnamefont {{Charisi}}}, \bibinfo {author}
  {\bibfnamefont {S.}~\bibnamefont {{Chatterjee}}}, \bibinfo {author}
  {\bibfnamefont {T.}~\bibnamefont {{Cohen}}}, \bibinfo {author} {\bibfnamefont
  {J.~M.}\ \bibnamefont {{Cordes}}}, \bibinfo {author} {\bibfnamefont {N.~J.}\
  \bibnamefont {{Cornish}}}, \bibinfo {author} {\bibfnamefont {F.}~\bibnamefont
  {{Crawford}}}, \bibinfo {author} {\bibfnamefont {H.~T.}\ \bibnamefont
  {{Cromartie}}}, \bibinfo {author} {\bibfnamefont {K.}~\bibnamefont
  {{Crowter}}}, \bibinfo {author} {\bibfnamefont {M.~E.}\ \bibnamefont
  {{Decesar}}}, \bibinfo {author} {\bibfnamefont {P.~B.}\ \bibnamefont
  {{Demorest}}}, \bibinfo {author} {\bibfnamefont {T.}~\bibnamefont {{Dolch}}},
  \bibinfo {author} {\bibfnamefont {B.}~\bibnamefont {{Drachler}}}, \bibinfo
  {author} {\bibfnamefont {E.~C.}\ \bibnamefont {{Ferrara}}}, \bibinfo {author}
  {\bibfnamefont {W.}~\bibnamefont {{Fiore}}}, \bibinfo {author} {\bibfnamefont
  {E.}~\bibnamefont {{Fonseca}}}, \bibinfo {author} {\bibfnamefont {G.~E.}\
  \bibnamefont {{Freedman}}}, \bibinfo {author} {\bibfnamefont
  {N.}~\bibnamefont {{Garver-Daniels}}}, \bibinfo {author} {\bibfnamefont
  {P.~A.}\ \bibnamefont {{Gentile}}}, \bibinfo {author} {\bibfnamefont
  {J.}~\bibnamefont {{Glaser}}}, \bibinfo {author} {\bibfnamefont {D.~C.}\
  \bibnamefont {{Good}}}, \bibinfo {author} {\bibfnamefont {K.}~\bibnamefont
  {{G{\"u}ltekin}}}, \bibinfo {author} {\bibfnamefont {J.~S.}\ \bibnamefont
  {{Hazboun}}}, \bibinfo {author} {\bibfnamefont {R.~J.}\ \bibnamefont
  {{Jennings}}}, \bibinfo {author} {\bibfnamefont {C.}~\bibnamefont
  {{Jessup}}}, \bibinfo {author} {\bibfnamefont {A.~D.}\ \bibnamefont
  {{Johnson}}}, \bibinfo {author} {\bibfnamefont {M.~L.}\ \bibnamefont
  {{Jones}}}, \bibinfo {author} {\bibfnamefont {A.~R.}\ \bibnamefont
  {{Kaiser}}}, \bibinfo {author} {\bibfnamefont {D.~L.}\ \bibnamefont
  {{Kaplan}}}, \bibinfo {author} {\bibfnamefont {L.~Z.}\ \bibnamefont
  {{Kelley}}}, \bibinfo {author} {\bibfnamefont {M.}~\bibnamefont {{Kerr}}},
  \bibinfo {author} {\bibfnamefont {J.~S.}\ \bibnamefont {{Key}}}, \bibinfo
  {author} {\bibfnamefont {A.}~\bibnamefont {{Kuske}}}, \bibinfo {author}
  {\bibfnamefont {N.}~\bibnamefont {{Laal}}}, \bibinfo {author} {\bibfnamefont
  {M.~T.}\ \bibnamefont {{Lam}}}, \bibinfo {author} {\bibfnamefont {W.~G.}\
  \bibnamefont {{Lamb}}}, \bibinfo {author} {\bibfnamefont {T.~J.~W.}\
  \bibnamefont {{Lazio}}}, \bibinfo {author} {\bibfnamefont {N.}~\bibnamefont
  {{Lewandowska}}}, \bibinfo {author} {\bibfnamefont {Y.}~\bibnamefont
  {{Lin}}}, \bibinfo {author} {\bibfnamefont {T.}~\bibnamefont {{Liu}}},
  \bibinfo {author} {\bibfnamefont {D.~R.}\ \bibnamefont {{Lorimer}}}, \bibinfo
  {author} {\bibfnamefont {J.}~\bibnamefont {{Luo}}}, \bibinfo {author}
  {\bibfnamefont {R.~S.}\ \bibnamefont {{Lynch}}}, \bibinfo {author}
  {\bibfnamefont {C.-P.}\ \bibnamefont {{Ma}}}, \bibinfo {author}
  {\bibfnamefont {D.~R.}\ \bibnamefont {{Madison}}}, \bibinfo {author}
  {\bibfnamefont {K.}~\bibnamefont {{Maraccini}}}, \bibinfo {author}
  {\bibfnamefont {A.}~\bibnamefont {{McEwen}}}, \bibinfo {author}
  {\bibfnamefont {J.~W.}\ \bibnamefont {{McKee}}}, \bibinfo {author}
  {\bibfnamefont {M.~A.}\ \bibnamefont {{McLaughlin}}}, \bibinfo {author}
  {\bibfnamefont {N.}~\bibnamefont {{McMann}}}, \bibinfo {author}
  {\bibfnamefont {B.~W.}\ \bibnamefont {{Meyers}}}, \bibinfo {author}
  {\bibfnamefont {C.~M.~F.}\ \bibnamefont {{Mingarelli}}}, \bibinfo {author}
  {\bibfnamefont {A.}~\bibnamefont {{Mitridate}}}, \bibinfo {author}
  {\bibfnamefont {C.}~\bibnamefont {{Ng}}}, \bibinfo {author} {\bibfnamefont
  {D.~J.}\ \bibnamefont {{Nice}}}, \bibinfo {author} {\bibfnamefont {S.~K.}\
  \bibnamefont {{Ocker}}}, \bibinfo {author} {\bibfnamefont {K.~D.}\
  \bibnamefont {{Olum}}}, \bibinfo {author} {\bibfnamefont {E.}~\bibnamefont
  {{Panciu}}}, \bibinfo {author} {\bibfnamefont {T.~T.}\ \bibnamefont
  {{Pennucci}}}, \bibinfo {author} {\bibfnamefont {B.~B.~P.}\ \bibnamefont
  {{Perera}}}, \bibinfo {author} {\bibfnamefont {N.~S.}\ \bibnamefont {{Pol}}},
  \bibinfo {author} {\bibfnamefont {H.~A.}\ \bibnamefont {{Radovan}}}, \bibinfo
  {author} {\bibfnamefont {S.~M.}\ \bibnamefont {{Ransom}}}, \bibinfo {author}
  {\bibfnamefont {P.~S.}\ \bibnamefont {{Ray}}}, \bibinfo {author}
  {\bibfnamefont {J.~D.}\ \bibnamefont {{Romano}}}, \bibinfo {author}
  {\bibfnamefont {L.}~\bibnamefont {{Salo}}}, \bibinfo {author} {\bibfnamefont
  {S.~C.}\ \bibnamefont {{Sardesai}}}, \bibinfo {author} {\bibfnamefont
  {C.}~\bibnamefont {{Schmiedekamp}}}, \bibinfo {author} {\bibfnamefont
  {A.}~\bibnamefont {{Schmiedekamp}}}, \bibinfo {author} {\bibfnamefont
  {K.}~\bibnamefont {{Schmitz}}}, \bibinfo {author} {\bibfnamefont {B.~J.}\
  \bibnamefont {{Shapiro-Albert}}}, \bibinfo {author} {\bibfnamefont
  {X.}~\bibnamefont {{Siemens}}}, \bibinfo {author} {\bibfnamefont
  {J.}~\bibnamefont {{Simon}}}, \bibinfo {author} {\bibfnamefont {M.~S.}\
  \bibnamefont {{Siwek}}}, \bibinfo {author} {\bibfnamefont {I.~H.}\
  \bibnamefont {{Stairs}}}, \bibinfo {author} {\bibfnamefont {D.~R.}\
  \bibnamefont {{Stinebring}}}, \bibinfo {author} {\bibfnamefont
  {K.}~\bibnamefont {{Stovall}}}, \bibinfo {author} {\bibfnamefont
  {A.}~\bibnamefont {{Susobhanan}}}, \bibinfo {author} {\bibfnamefont {J.~K.}\
  \bibnamefont {{Swiggum}}}, \bibinfo {author} {\bibfnamefont {S.~R.}\
  \bibnamefont {{Taylor}}}, \bibinfo {author} {\bibfnamefont {J.~E.}\
  \bibnamefont {{Turner}}}, \bibinfo {author} {\bibfnamefont {C.}~\bibnamefont
  {{Unal}}}, \bibinfo {author} {\bibfnamefont {M.}~\bibnamefont
  {{Vallisneri}}}, \bibinfo {author} {\bibfnamefont {S.~J.}\ \bibnamefont
  {{Vigeland}}}, \bibinfo {author} {\bibfnamefont {H.~M.}\ \bibnamefont
  {{Wahl}}}, \bibinfo {author} {\bibfnamefont {Q.}~\bibnamefont {{Wang}}},
  \bibinfo {author} {\bibfnamefont {C.~A.}\ \bibnamefont {{Witt}}}, \bibinfo
  {author} {\bibfnamefont {O.}~\bibnamefont {{Young}}},\ and\ \bibinfo {author}
  {\bibnamefont {{Nanograv Collaboration}}},\ }\bibfield  {title} {\bibinfo
  {title} {{The NANOGrav 15 yr Data Set: Observations and Timing of 68
  Millisecond Pulsars}},\ }\href {https://doi.org/10.3847/2041-8213/acda9a}
  {\bibfield  {journal} {\bibinfo  {journal} {\apjl}\ }\textbf {\bibinfo
  {volume} {951}},\ \bibinfo {eid} {L9} (\bibinfo {year} {2023})},\ \Eprint
  {https://arxiv.org/abs/2306.16217} {arXiv:2306.16217 [astro-ph.HE]}
  \BibitemShut {NoStop}%
\bibitem [{\citenamefont {{EPTA Collaboration}}\ \emph
  {et~al.}(2023)\citenamefont {{EPTA Collaboration}}, \citenamefont {{InPTA
  Collaboration}}, \citenamefont {{Antoniadis}}, \citenamefont {{Arumugam}},
  \citenamefont {{Arumugam}}, \citenamefont {{Babak}}, \citenamefont
  {{Bagchi}}, \citenamefont {{Bak Nielsen}}, \citenamefont {{Bassa}},
  \citenamefont {{Bathula}}, \citenamefont {{Berthereau}}, \citenamefont
  {{Bonetti}}, \citenamefont {{Bortolas}}, \citenamefont {{Brook}},
  \citenamefont {{Burgay}}, \citenamefont {{Caballero}}, \citenamefont
  {{Chalumeau}}, \citenamefont {{Champion}}, \citenamefont {{Chanlaridis}},
  \citenamefont {{Chen}}, \citenamefont {{Cognard}}, \citenamefont
  {{Dandapat}}, \citenamefont {{Deb}}, \citenamefont {{Desai}}, \citenamefont
  {{Desvignes}}, \citenamefont {{Dhanda-Batra}}, \citenamefont {{Dwivedi}},
  \citenamefont {{Falxa}}, \citenamefont {{Ferdman}}, \citenamefont
  {{Franchini}}, \citenamefont {{Gair}}, \citenamefont {{Goncharov}},
  \citenamefont {{Gopakumar}}, \citenamefont {{Graikou}}, \citenamefont
  {{Grie{\ss}meier}}, \citenamefont {{Guillemot}}, \citenamefont {{Guo}},
  \citenamefont {{Gupta}}, \citenamefont {{Hisano}}, \citenamefont {{Hu}},
  \citenamefont {{Iraci}}, \citenamefont {{Izquierdo-Villalba}}, \citenamefont
  {{Jang}}, \citenamefont {{Jawor}}, \citenamefont {{Janssen}}, \citenamefont
  {{Jessner}}, \citenamefont {{Joshi}}, \citenamefont {{Kareem}}, \citenamefont
  {{Karuppusamy}}, \citenamefont {{Keane}}, \citenamefont {{Keith}},
  \citenamefont {{Kharbanda}}, \citenamefont {{Kikunaga}}, \citenamefont
  {{Kolhe}}, \citenamefont {{Kramer}}, \citenamefont {{Krishnakumar}},
  \citenamefont {{Lackeos}}, \citenamefont {{Lee}}, \citenamefont {{Liu}},
  \citenamefont {{Liu}}, \citenamefont {{Lyne}}, \citenamefont {{McKee}},
  \citenamefont {{Maan}}, \citenamefont {{Main}}, \citenamefont {{Mickaliger}},
  \citenamefont {{Ni{\c{t}}u}}, \citenamefont {{Nobleson}}, \citenamefont
  {{Paladi}}, \citenamefont {{Parthasarathy}}, \citenamefont {{Perera}},
  \citenamefont {{Perrodin}}, \citenamefont {{Petiteau}}, \citenamefont
  {{Porayko}}, \citenamefont {{Possenti}}, \citenamefont {{Prabu}},
  \citenamefont {{Quelquejay Leclere}}, \citenamefont {{Rana}}, \citenamefont
  {{Samajdar}}, \citenamefont {{Sanidas}}, \citenamefont {{Sesana}},
  \citenamefont {{Shaifullah}}, \citenamefont {{Singha}}, \citenamefont
  {{Speri}}, \citenamefont {{Spiewak}}, \citenamefont {{Srivastava}},
  \citenamefont {{Stappers}}, \citenamefont {{Surnis}}, \citenamefont
  {{Susarla}}, \citenamefont {{Susobhanan}}, \citenamefont {{Takahashi}},
  \citenamefont {{Tarafdar}}, \citenamefont {{Theureau}}, \citenamefont
  {{Tiburzi}}, \citenamefont {{van der Wateren}}, \citenamefont {{Vecchio}},
  \citenamefont {{Venkatraman Krishnan}}, \citenamefont {{Verbiest}},
  \citenamefont {{Wang}}, \citenamefont {{Wang}},\ and\ \citenamefont
  {{Wu}}}]{2023A&A...678A..50E}%
  \BibitemOpen
  \bibfield  {author} {\bibinfo {author} {\bibnamefont {{EPTA Collaboration}}},
  \bibinfo {author} {\bibnamefont {{InPTA Collaboration}}}, \bibinfo {author}
  {\bibfnamefont {J.}~\bibnamefont {{Antoniadis}}}, \bibinfo {author}
  {\bibfnamefont {P.}~\bibnamefont {{Arumugam}}}, \bibinfo {author}
  {\bibfnamefont {S.}~\bibnamefont {{Arumugam}}}, \bibinfo {author}
  {\bibfnamefont {S.}~\bibnamefont {{Babak}}}, \bibinfo {author} {\bibfnamefont
  {M.}~\bibnamefont {{Bagchi}}}, \bibinfo {author} {\bibfnamefont {A.~S.}\
  \bibnamefont {{Bak Nielsen}}}, \bibinfo {author} {\bibfnamefont {C.~G.}\
  \bibnamefont {{Bassa}}}, \bibinfo {author} {\bibfnamefont {A.}~\bibnamefont
  {{Bathula}}}, \bibinfo {author} {\bibfnamefont {A.}~\bibnamefont
  {{Berthereau}}}, \bibinfo {author} {\bibfnamefont {M.}~\bibnamefont
  {{Bonetti}}}, \bibinfo {author} {\bibfnamefont {E.}~\bibnamefont
  {{Bortolas}}}, \bibinfo {author} {\bibfnamefont {P.~R.}\ \bibnamefont
  {{Brook}}}, \bibinfo {author} {\bibfnamefont {M.}~\bibnamefont {{Burgay}}},
  \bibinfo {author} {\bibfnamefont {R.~N.}\ \bibnamefont {{Caballero}}},
  \bibinfo {author} {\bibfnamefont {A.}~\bibnamefont {{Chalumeau}}}, \bibinfo
  {author} {\bibfnamefont {D.~J.}\ \bibnamefont {{Champion}}}, \bibinfo
  {author} {\bibfnamefont {S.}~\bibnamefont {{Chanlaridis}}}, \bibinfo {author}
  {\bibfnamefont {S.}~\bibnamefont {{Chen}}}, \bibinfo {author} {\bibfnamefont
  {I.}~\bibnamefont {{Cognard}}}, \bibinfo {author} {\bibfnamefont
  {S.}~\bibnamefont {{Dandapat}}}, \bibinfo {author} {\bibfnamefont
  {D.}~\bibnamefont {{Deb}}}, \bibinfo {author} {\bibfnamefont
  {S.}~\bibnamefont {{Desai}}}, \bibinfo {author} {\bibfnamefont
  {G.}~\bibnamefont {{Desvignes}}}, \bibinfo {author} {\bibfnamefont
  {N.}~\bibnamefont {{Dhanda-Batra}}}, \bibinfo {author} {\bibfnamefont
  {C.}~\bibnamefont {{Dwivedi}}}, \bibinfo {author} {\bibfnamefont
  {M.}~\bibnamefont {{Falxa}}}, \bibinfo {author} {\bibfnamefont {R.~D.}\
  \bibnamefont {{Ferdman}}}, \bibinfo {author} {\bibfnamefont {A.}~\bibnamefont
  {{Franchini}}}, \bibinfo {author} {\bibfnamefont {J.~R.}\ \bibnamefont
  {{Gair}}}, \bibinfo {author} {\bibfnamefont {B.}~\bibnamefont {{Goncharov}}},
  \bibinfo {author} {\bibfnamefont {A.}~\bibnamefont {{Gopakumar}}}, \bibinfo
  {author} {\bibfnamefont {E.}~\bibnamefont {{Graikou}}}, \bibinfo {author}
  {\bibfnamefont {J.~M.}\ \bibnamefont {{Grie{\ss}meier}}}, \bibinfo {author}
  {\bibfnamefont {L.}~\bibnamefont {{Guillemot}}}, \bibinfo {author}
  {\bibfnamefont {Y.~J.}\ \bibnamefont {{Guo}}}, \bibinfo {author}
  {\bibfnamefont {Y.}~\bibnamefont {{Gupta}}}, \bibinfo {author} {\bibfnamefont
  {S.}~\bibnamefont {{Hisano}}}, \bibinfo {author} {\bibfnamefont
  {H.}~\bibnamefont {{Hu}}}, \bibinfo {author} {\bibfnamefont {F.}~\bibnamefont
  {{Iraci}}}, \bibinfo {author} {\bibfnamefont {D.}~\bibnamefont
  {{Izquierdo-Villalba}}}, \bibinfo {author} {\bibfnamefont {J.}~\bibnamefont
  {{Jang}}}, \bibinfo {author} {\bibfnamefont {J.}~\bibnamefont {{Jawor}}},
  \bibinfo {author} {\bibfnamefont {G.~H.}\ \bibnamefont {{Janssen}}}, \bibinfo
  {author} {\bibfnamefont {A.}~\bibnamefont {{Jessner}}}, \bibinfo {author}
  {\bibfnamefont {B.~C.}\ \bibnamefont {{Joshi}}}, \bibinfo {author}
  {\bibfnamefont {F.}~\bibnamefont {{Kareem}}}, \bibinfo {author}
  {\bibfnamefont {R.}~\bibnamefont {{Karuppusamy}}}, \bibinfo {author}
  {\bibfnamefont {E.~F.}\ \bibnamefont {{Keane}}}, \bibinfo {author}
  {\bibfnamefont {M.~J.}\ \bibnamefont {{Keith}}}, \bibinfo {author}
  {\bibfnamefont {D.}~\bibnamefont {{Kharbanda}}}, \bibinfo {author}
  {\bibfnamefont {T.}~\bibnamefont {{Kikunaga}}}, \bibinfo {author}
  {\bibfnamefont {N.}~\bibnamefont {{Kolhe}}}, \bibinfo {author} {\bibfnamefont
  {M.}~\bibnamefont {{Kramer}}}, \bibinfo {author} {\bibfnamefont {M.~A.}\
  \bibnamefont {{Krishnakumar}}}, \bibinfo {author} {\bibfnamefont
  {K.}~\bibnamefont {{Lackeos}}}, \bibinfo {author} {\bibfnamefont {K.~J.}\
  \bibnamefont {{Lee}}}, \bibinfo {author} {\bibfnamefont {K.}~\bibnamefont
  {{Liu}}}, \bibinfo {author} {\bibfnamefont {Y.}~\bibnamefont {{Liu}}},
  \bibinfo {author} {\bibfnamefont {A.~G.}\ \bibnamefont {{Lyne}}}, \bibinfo
  {author} {\bibfnamefont {J.~W.}\ \bibnamefont {{McKee}}}, \bibinfo {author}
  {\bibfnamefont {Y.}~\bibnamefont {{Maan}}}, \bibinfo {author} {\bibfnamefont
  {R.~A.}\ \bibnamefont {{Main}}}, \bibinfo {author} {\bibfnamefont {M.~B.}\
  \bibnamefont {{Mickaliger}}}, \bibinfo {author} {\bibfnamefont {I.~C.}\
  \bibnamefont {{Ni{\c{t}}u}}}, \bibinfo {author} {\bibfnamefont
  {K.}~\bibnamefont {{Nobleson}}}, \bibinfo {author} {\bibfnamefont {A.~K.}\
  \bibnamefont {{Paladi}}}, \bibinfo {author} {\bibfnamefont {A.}~\bibnamefont
  {{Parthasarathy}}}, \bibinfo {author} {\bibfnamefont {B.~B.~P.}\ \bibnamefont
  {{Perera}}}, \bibinfo {author} {\bibfnamefont {D.}~\bibnamefont
  {{Perrodin}}}, \bibinfo {author} {\bibfnamefont {A.}~\bibnamefont
  {{Petiteau}}}, \bibinfo {author} {\bibfnamefont {N.~K.}\ \bibnamefont
  {{Porayko}}}, \bibinfo {author} {\bibfnamefont {A.}~\bibnamefont
  {{Possenti}}}, \bibinfo {author} {\bibfnamefont {T.}~\bibnamefont {{Prabu}}},
  \bibinfo {author} {\bibfnamefont {H.}~\bibnamefont {{Quelquejay Leclere}}},
  \bibinfo {author} {\bibfnamefont {P.}~\bibnamefont {{Rana}}}, \bibinfo
  {author} {\bibfnamefont {A.}~\bibnamefont {{Samajdar}}}, \bibinfo {author}
  {\bibfnamefont {S.~A.}\ \bibnamefont {{Sanidas}}}, \bibinfo {author}
  {\bibfnamefont {A.}~\bibnamefont {{Sesana}}}, \bibinfo {author}
  {\bibfnamefont {G.}~\bibnamefont {{Shaifullah}}}, \bibinfo {author}
  {\bibfnamefont {J.}~\bibnamefont {{Singha}}}, \bibinfo {author}
  {\bibfnamefont {L.}~\bibnamefont {{Speri}}}, \bibinfo {author} {\bibfnamefont
  {R.}~\bibnamefont {{Spiewak}}}, \bibinfo {author} {\bibfnamefont
  {A.}~\bibnamefont {{Srivastava}}}, \bibinfo {author} {\bibfnamefont {B.~W.}\
  \bibnamefont {{Stappers}}}, \bibinfo {author} {\bibfnamefont
  {M.}~\bibnamefont {{Surnis}}}, \bibinfo {author} {\bibfnamefont {S.~C.}\
  \bibnamefont {{Susarla}}}, \bibinfo {author} {\bibfnamefont {A.}~\bibnamefont
  {{Susobhanan}}}, \bibinfo {author} {\bibfnamefont {K.}~\bibnamefont
  {{Takahashi}}}, \bibinfo {author} {\bibfnamefont {P.}~\bibnamefont
  {{Tarafdar}}}, \bibinfo {author} {\bibfnamefont {G.}~\bibnamefont
  {{Theureau}}}, \bibinfo {author} {\bibfnamefont {C.}~\bibnamefont
  {{Tiburzi}}}, \bibinfo {author} {\bibfnamefont {E.}~\bibnamefont {{van der
  Wateren}}}, \bibinfo {author} {\bibfnamefont {A.}~\bibnamefont {{Vecchio}}},
  \bibinfo {author} {\bibfnamefont {V.}~\bibnamefont {{Venkatraman Krishnan}}},
  \bibinfo {author} {\bibfnamefont {J.~P.~W.}\ \bibnamefont {{Verbiest}}},
  \bibinfo {author} {\bibfnamefont {J.}~\bibnamefont {{Wang}}}, \bibinfo
  {author} {\bibfnamefont {L.}~\bibnamefont {{Wang}}},\ and\ \bibinfo {author}
  {\bibfnamefont {Z.}~\bibnamefont {{Wu}}},\ }\bibfield  {title} {\bibinfo
  {title} {{The second data release from the European Pulsar Timing Array. III.
  Search for gravitational wave signals}},\ }\href
  {https://doi.org/10.1051/0004-6361/202346844} {\bibfield  {journal} {\bibinfo
   {journal} {\aap}\ }\textbf {\bibinfo {volume} {678}},\ \bibinfo {eid} {A50}
  (\bibinfo {year} {2023})},\ \Eprint {https://arxiv.org/abs/2306.16214}
  {arXiv:2306.16214 [astro-ph.HE]} \BibitemShut {NoStop}%
\bibitem [{\citenamefont {{Reardon}}\ \emph {et~al.}(2023)\citenamefont
  {{Reardon}}, \citenamefont {{Zic}}, \citenamefont {{Shannon}}, \citenamefont
  {{Hobbs}}, \citenamefont {{Bailes}}, \citenamefont {{Di Marco}},
  \citenamefont {{Kapur}}, \citenamefont {{Rogers}}, \citenamefont {{Thrane}},
  \citenamefont {{Askew}}, \citenamefont {{Bhat}}, \citenamefont {{Cameron}},
  \citenamefont {{Cury{\l}o}}, \citenamefont {{Coles}}, \citenamefont {{Dai}},
  \citenamefont {{Goncharov}}, \citenamefont {{Kerr}}, \citenamefont
  {{Kulkarni}}, \citenamefont {{Levin}}, \citenamefont {{Lower}}, \citenamefont
  {{Manchester}}, \citenamefont {{Mandow}}, \citenamefont {{Miles}},
  \citenamefont {{Nathan}}, \citenamefont {{Os{\l}owski}}, \citenamefont
  {{Russell}}, \citenamefont {{Spiewak}}, \citenamefont {{Zhang}},\ and\
  \citenamefont {{Zhu}}}]{2023ApJ...951L...6R}%
  \BibitemOpen
  \bibfield  {author} {\bibinfo {author} {\bibfnamefont {D.~J.}\ \bibnamefont
  {{Reardon}}}, \bibinfo {author} {\bibfnamefont {A.}~\bibnamefont {{Zic}}},
  \bibinfo {author} {\bibfnamefont {R.~M.}\ \bibnamefont {{Shannon}}}, \bibinfo
  {author} {\bibfnamefont {G.~B.}\ \bibnamefont {{Hobbs}}}, \bibinfo {author}
  {\bibfnamefont {M.}~\bibnamefont {{Bailes}}}, \bibinfo {author}
  {\bibfnamefont {V.}~\bibnamefont {{Di Marco}}}, \bibinfo {author}
  {\bibfnamefont {A.}~\bibnamefont {{Kapur}}}, \bibinfo {author} {\bibfnamefont
  {A.~F.}\ \bibnamefont {{Rogers}}}, \bibinfo {author} {\bibfnamefont
  {E.}~\bibnamefont {{Thrane}}}, \bibinfo {author} {\bibfnamefont
  {J.}~\bibnamefont {{Askew}}}, \bibinfo {author} {\bibfnamefont {N.~D.~R.}\
  \bibnamefont {{Bhat}}}, \bibinfo {author} {\bibfnamefont {A.}~\bibnamefont
  {{Cameron}}}, \bibinfo {author} {\bibfnamefont {M.}~\bibnamefont
  {{Cury{\l}o}}}, \bibinfo {author} {\bibfnamefont {W.~A.}\ \bibnamefont
  {{Coles}}}, \bibinfo {author} {\bibfnamefont {S.}~\bibnamefont {{Dai}}},
  \bibinfo {author} {\bibfnamefont {B.}~\bibnamefont {{Goncharov}}}, \bibinfo
  {author} {\bibfnamefont {M.}~\bibnamefont {{Kerr}}}, \bibinfo {author}
  {\bibfnamefont {A.}~\bibnamefont {{Kulkarni}}}, \bibinfo {author}
  {\bibfnamefont {Y.}~\bibnamefont {{Levin}}}, \bibinfo {author} {\bibfnamefont
  {M.~E.}\ \bibnamefont {{Lower}}}, \bibinfo {author} {\bibfnamefont {R.~N.}\
  \bibnamefont {{Manchester}}}, \bibinfo {author} {\bibfnamefont
  {R.}~\bibnamefont {{Mandow}}}, \bibinfo {author} {\bibfnamefont {M.~T.}\
  \bibnamefont {{Miles}}}, \bibinfo {author} {\bibfnamefont {R.~S.}\
  \bibnamefont {{Nathan}}}, \bibinfo {author} {\bibfnamefont {S.}~\bibnamefont
  {{Os{\l}owski}}}, \bibinfo {author} {\bibfnamefont {C.~J.}\ \bibnamefont
  {{Russell}}}, \bibinfo {author} {\bibfnamefont {R.}~\bibnamefont
  {{Spiewak}}}, \bibinfo {author} {\bibfnamefont {S.}~\bibnamefont {{Zhang}}},\
  and\ \bibinfo {author} {\bibfnamefont {X.-J.}\ \bibnamefont {{Zhu}}},\
  }\bibfield  {title} {\bibinfo {title} {{Search for an Isotropic
  Gravitational-wave Background with the Parkes Pulsar Timing Array}},\ }\href
  {https://doi.org/10.3847/2041-8213/acdd02} {\bibfield  {journal} {\bibinfo
  {journal} {\apjl}\ }\textbf {\bibinfo {volume} {951}},\ \bibinfo {eid} {L6}
  (\bibinfo {year} {2023})},\ \Eprint {https://arxiv.org/abs/2306.16215}
  {arXiv:2306.16215 [astro-ph.HE]} \BibitemShut {NoStop}%
\bibitem [{\citenamefont {{Xu}}\ \emph {et~al.}(2023)\citenamefont {{Xu}},
  \citenamefont {{Chen}}, \citenamefont {{Guo}}, \citenamefont {{Jiang}},
  \citenamefont {{Wang}}, \citenamefont {{Xu}}, \citenamefont {{Xue}},
  \citenamefont {{Caballero}}, \citenamefont {{Yuan}}, \citenamefont {{Xu}},
  \citenamefont {{Wang}}, \citenamefont {{Hao}}, \citenamefont {{Luo}},
  \citenamefont {{Lee}}, \citenamefont {{Han}}, \citenamefont {{Jiang}},
  \citenamefont {{Shen}}, \citenamefont {{Wang}}, \citenamefont {{Wang}},
  \citenamefont {{Xu}}, \citenamefont {{Wu}}, \citenamefont {{Manchester}},
  \citenamefont {{Qian}}, \citenamefont {{Guan}}, \citenamefont {{Huang}},
  \citenamefont {{Sun}},\ and\ \citenamefont {{Zhu}}}]{2023RAA....23g5024X}%
  \BibitemOpen
  \bibfield  {author} {\bibinfo {author} {\bibfnamefont {H.}~\bibnamefont
  {{Xu}}}, \bibinfo {author} {\bibfnamefont {S.}~\bibnamefont {{Chen}}},
  \bibinfo {author} {\bibfnamefont {Y.}~\bibnamefont {{Guo}}}, \bibinfo
  {author} {\bibfnamefont {J.}~\bibnamefont {{Jiang}}}, \bibinfo {author}
  {\bibfnamefont {B.}~\bibnamefont {{Wang}}}, \bibinfo {author} {\bibfnamefont
  {J.}~\bibnamefont {{Xu}}}, \bibinfo {author} {\bibfnamefont {Z.}~\bibnamefont
  {{Xue}}}, \bibinfo {author} {\bibfnamefont {R.~N.}\ \bibnamefont
  {{Caballero}}}, \bibinfo {author} {\bibfnamefont {J.}~\bibnamefont {{Yuan}}},
  \bibinfo {author} {\bibfnamefont {Y.}~\bibnamefont {{Xu}}}, \bibinfo {author}
  {\bibfnamefont {J.}~\bibnamefont {{Wang}}}, \bibinfo {author} {\bibfnamefont
  {L.}~\bibnamefont {{Hao}}}, \bibinfo {author} {\bibfnamefont
  {J.}~\bibnamefont {{Luo}}}, \bibinfo {author} {\bibfnamefont
  {K.}~\bibnamefont {{Lee}}}, \bibinfo {author} {\bibfnamefont
  {J.}~\bibnamefont {{Han}}}, \bibinfo {author} {\bibfnamefont
  {P.}~\bibnamefont {{Jiang}}}, \bibinfo {author} {\bibfnamefont
  {Z.}~\bibnamefont {{Shen}}}, \bibinfo {author} {\bibfnamefont
  {M.}~\bibnamefont {{Wang}}}, \bibinfo {author} {\bibfnamefont
  {N.}~\bibnamefont {{Wang}}}, \bibinfo {author} {\bibfnamefont
  {R.}~\bibnamefont {{Xu}}}, \bibinfo {author} {\bibfnamefont {X.}~\bibnamefont
  {{Wu}}}, \bibinfo {author} {\bibfnamefont {R.}~\bibnamefont {{Manchester}}},
  \bibinfo {author} {\bibfnamefont {L.}~\bibnamefont {{Qian}}}, \bibinfo
  {author} {\bibfnamefont {X.}~\bibnamefont {{Guan}}}, \bibinfo {author}
  {\bibfnamefont {M.}~\bibnamefont {{Huang}}}, \bibinfo {author} {\bibfnamefont
  {C.}~\bibnamefont {{Sun}}},\ and\ \bibinfo {author} {\bibfnamefont
  {Y.}~\bibnamefont {{Zhu}}},\ }\bibfield  {title} {\bibinfo {title}
  {{Searching for the Nano-Hertz Stochastic Gravitational Wave Background with
  the Chinese Pulsar Timing Array Data Release I}},\ }\href
  {https://doi.org/10.1088/1674-4527/acdfa5} {\bibfield  {journal} {\bibinfo
  {journal} {Research in Astronomy and Astrophysics}\ }\textbf {\bibinfo
  {volume} {23}},\ \bibinfo {eid} {075024} (\bibinfo {year} {2023})},\ \Eprint
  {https://arxiv.org/abs/2306.16216} {arXiv:2306.16216 [astro-ph.HE]}
  \BibitemShut {NoStop}%
\bibitem [{\citenamefont {{Miles}}\ \emph {et~al.}(2025)\citenamefont
  {{Miles}}, \citenamefont {{Shannon}}, \citenamefont {{Reardon}},
  \citenamefont {{Bailes}}, \citenamefont {{Champion}}, \citenamefont
  {{Geyer}}, \citenamefont {{Gitika}}, \citenamefont {{Grunthal}},
  \citenamefont {{Keith}}, \citenamefont {{Kramer}}, \citenamefont
  {{Kulkarni}}, \citenamefont {{Nathan}}, \citenamefont {{Parthasarathy}},
  \citenamefont {{Singha}}, \citenamefont {{Theureau}}, \citenamefont
  {{Thrane}}, \citenamefont {{Abbate}}, \citenamefont {{Buchner}},
  \citenamefont {{Cameron}}, \citenamefont {{Camilo}}, \citenamefont
  {{Moreschi}}, \citenamefont {{Shaifullah}}, \citenamefont {{Shamohammadi}},
  \citenamefont {{Possenti}},\ and\ \citenamefont
  {{Krishnan}}}]{2025MNRAS.536.1489M}%
  \BibitemOpen
  \bibfield  {author} {\bibinfo {author} {\bibfnamefont {M.~T.}\ \bibnamefont
  {{Miles}}}, \bibinfo {author} {\bibfnamefont {R.~M.}\ \bibnamefont
  {{Shannon}}}, \bibinfo {author} {\bibfnamefont {D.~J.}\ \bibnamefont
  {{Reardon}}}, \bibinfo {author} {\bibfnamefont {M.}~\bibnamefont {{Bailes}}},
  \bibinfo {author} {\bibfnamefont {D.~J.}\ \bibnamefont {{Champion}}},
  \bibinfo {author} {\bibfnamefont {M.}~\bibnamefont {{Geyer}}}, \bibinfo
  {author} {\bibfnamefont {P.}~\bibnamefont {{Gitika}}}, \bibinfo {author}
  {\bibfnamefont {K.}~\bibnamefont {{Grunthal}}}, \bibinfo {author}
  {\bibfnamefont {M.~J.}\ \bibnamefont {{Keith}}}, \bibinfo {author}
  {\bibfnamefont {M.}~\bibnamefont {{Kramer}}}, \bibinfo {author}
  {\bibfnamefont {A.~D.}\ \bibnamefont {{Kulkarni}}}, \bibinfo {author}
  {\bibfnamefont {R.~S.}\ \bibnamefont {{Nathan}}}, \bibinfo {author}
  {\bibfnamefont {A.}~\bibnamefont {{Parthasarathy}}}, \bibinfo {author}
  {\bibfnamefont {J.}~\bibnamefont {{Singha}}}, \bibinfo {author}
  {\bibfnamefont {G.}~\bibnamefont {{Theureau}}}, \bibinfo {author}
  {\bibfnamefont {E.}~\bibnamefont {{Thrane}}}, \bibinfo {author}
  {\bibfnamefont {F.}~\bibnamefont {{Abbate}}}, \bibinfo {author}
  {\bibfnamefont {S.}~\bibnamefont {{Buchner}}}, \bibinfo {author}
  {\bibfnamefont {A.~D.}\ \bibnamefont {{Cameron}}}, \bibinfo {author}
  {\bibfnamefont {F.}~\bibnamefont {{Camilo}}}, \bibinfo {author}
  {\bibfnamefont {B.~E.}\ \bibnamefont {{Moreschi}}}, \bibinfo {author}
  {\bibfnamefont {G.}~\bibnamefont {{Shaifullah}}}, \bibinfo {author}
  {\bibfnamefont {M.}~\bibnamefont {{Shamohammadi}}}, \bibinfo {author}
  {\bibfnamefont {A.}~\bibnamefont {{Possenti}}},\ and\ \bibinfo {author}
  {\bibfnamefont {V.~V.}\ \bibnamefont {{Krishnan}}},\ }\bibfield  {title}
  {\bibinfo {title} {{The MeerKAT Pulsar Timing Array: the first search for
  gravitational waves with the MeerKAT radio telescope}},\ }\href
  {https://doi.org/10.1093/mnras/stae2571} {\bibfield  {journal} {\bibinfo
  {journal} {\mnras}\ }\textbf {\bibinfo {volume} {536}},\ \bibinfo {pages}
  {1489} (\bibinfo {year} {2025})},\ \Eprint {https://arxiv.org/abs/2412.01153}
  {arXiv:2412.01153 [astro-ph.HE]} \BibitemShut {NoStop}%
\bibitem [{\citenamefont {{Lyne}}\ \emph {et~al.}(1987)\citenamefont {{Lyne}},
  \citenamefont {{Brinklow}}, \citenamefont {{Middleditch}}, \citenamefont
  {{Kulkarni}},\ and\ \citenamefont {{Backer}}}]{1987Natur.328..399L}%
  \BibitemOpen
  \bibfield  {author} {\bibinfo {author} {\bibfnamefont {A.~G.}\ \bibnamefont
  {{Lyne}}}, \bibinfo {author} {\bibfnamefont {A.}~\bibnamefont {{Brinklow}}},
  \bibinfo {author} {\bibfnamefont {J.}~\bibnamefont {{Middleditch}}}, \bibinfo
  {author} {\bibfnamefont {S.~R.}\ \bibnamefont {{Kulkarni}}},\ and\ \bibinfo
  {author} {\bibfnamefont {D.~C.}\ \bibnamefont {{Backer}}},\ }\bibfield
  {title} {\bibinfo {title} {{The discovery of a millisecond pulsar in the
  globular cluster M28}},\ }\href {https://doi.org/10.1038/328399a0} {\bibfield
   {journal} {\bibinfo  {journal} {\nat}\ }\textbf {\bibinfo {volume} {328}},\
  \bibinfo {pages} {399} (\bibinfo {year} {1987})}\BibitemShut {NoStop}%
\bibitem [{Note1()}]{Note1}%
  \BibitemOpen
  \bibinfo {note} {Pulsars in GCs: \protect \url
  {https://www3.mpifr-bonn.mpg.de/staff/pfreire/GCpsr.html}}\BibitemShut
  {NoStop}%
\bibitem [{\citenamefont {{Yin}}\ \emph {et~al.}(2023)\citenamefont {{Yin}},
  \citenamefont {{Zhang}}, \citenamefont {{Li}}, \citenamefont {{Li}},
  \citenamefont {{Qian}},\ and\ \citenamefont {{Pan}}}]{2023RAA....23e5012Y}%
  \BibitemOpen
  \bibfield  {author} {\bibinfo {author} {\bibfnamefont {D.-J.}\ \bibnamefont
  {{Yin}}}, \bibinfo {author} {\bibfnamefont {L.-Y.}\ \bibnamefont {{Zhang}}},
  \bibinfo {author} {\bibfnamefont {B.-D.}\ \bibnamefont {{Li}}}, \bibinfo
  {author} {\bibfnamefont {M.-H.}\ \bibnamefont {{Li}}}, \bibinfo {author}
  {\bibfnamefont {L.}~\bibnamefont {{Qian}}},\ and\ \bibinfo {author}
  {\bibfnamefont {Z.}~\bibnamefont {{Pan}}},\ }\bibfield  {title} {\bibinfo
  {title} {{The Analyses of Globular Cluster Pulsars and Their Detection
  Efficiency}},\ }\href {https://doi.org/10.1088/1674-4527/acc37e} {\bibfield
  {journal} {\bibinfo  {journal} {Research in Astronomy and Astrophysics}\
  }\textbf {\bibinfo {volume} {23}},\ \bibinfo {eid} {055012} (\bibinfo {year}
  {2023})},\ \Eprint {https://arxiv.org/abs/2303.09710} {arXiv:2303.09710
  [astro-ph.HE]} \BibitemShut {NoStop}%
\bibitem [{\citenamefont {{Yin}}\ \emph {et~al.}(2024)\citenamefont {{Yin}},
  \citenamefont {{Zhang}}, \citenamefont {{Qian}}, \citenamefont {{Eatough}},
  \citenamefont {{Li}}, \citenamefont {{Lorimer}}, \citenamefont {{Dai}},
  \citenamefont {{Li}}, \citenamefont {{Zhang}}, \citenamefont {{Li}},
  \citenamefont {{Su}}, \citenamefont {{Wu}}, \citenamefont {{Pan}},
  \citenamefont {{Lian}}, \citenamefont {{Liu}}, \citenamefont {{Yan}},\ and\
  \citenamefont {{Pan}}}]{yin24}%
  \BibitemOpen
  \bibfield  {author} {\bibinfo {author} {\bibfnamefont {D.}~\bibnamefont
  {{Yin}}}, \bibinfo {author} {\bibfnamefont {L.-y.}\ \bibnamefont {{Zhang}}},
  \bibinfo {author} {\bibfnamefont {L.}~\bibnamefont {{Qian}}}, \bibinfo
  {author} {\bibfnamefont {R.~P.}\ \bibnamefont {{Eatough}}}, \bibinfo {author}
  {\bibfnamefont {B.}~\bibnamefont {{Li}}}, \bibinfo {author} {\bibfnamefont
  {D.~R.}\ \bibnamefont {{Lorimer}}}, \bibinfo {author} {\bibfnamefont
  {Y.}~\bibnamefont {{Dai}}}, \bibinfo {author} {\bibfnamefont
  {Y.}~\bibnamefont {{Li}}}, \bibinfo {author} {\bibfnamefont {X.}~\bibnamefont
  {{Zhang}}}, \bibinfo {author} {\bibfnamefont {M.}~\bibnamefont {{Li}}},
  \bibinfo {author} {\bibfnamefont {T.}~\bibnamefont {{Su}}}, \bibinfo {author}
  {\bibfnamefont {Y.}~\bibnamefont {{Wu}}}, \bibinfo {author} {\bibfnamefont
  {Y.}~\bibnamefont {{Pan}}}, \bibinfo {author} {\bibfnamefont
  {Y.}~\bibnamefont {{Lian}}}, \bibinfo {author} {\bibfnamefont
  {T.}~\bibnamefont {{Liu}}}, \bibinfo {author} {\bibfnamefont
  {Z.}~\bibnamefont {{Yan}}},\ and\ \bibinfo {author} {\bibfnamefont
  {Z.}~\bibnamefont {{Pan}}},\ }\bibfield  {title} {\bibinfo {title} {{FAST
  Discovery of Eight Isolated Millisecond Pulsars in NGC 6517}},\ }\href
  {https://doi.org/10.3847/2041-8213/ad534e} {\bibfield  {journal} {\bibinfo
  {journal} {\apjl}\ }\textbf {\bibinfo {volume} {969}},\ \bibinfo {eid} {L7}
  (\bibinfo {year} {2024})},\ \Eprint {https://arxiv.org/abs/2405.18228}
  {arXiv:2405.18228 [astro-ph.HE]} \BibitemShut {NoStop}%
\bibitem [{\citenamefont {{Wu}}\ \emph {et~al.}(2024)\citenamefont {{Wu}},
  \citenamefont {{Pan}}, \citenamefont {{Qian}}, \citenamefont {{Ransom}},
  \citenamefont {{Eatough}}, \citenamefont {{Wang}}, \citenamefont {{Freire}},
  \citenamefont {{Liu}}, \citenamefont {{Yan}}, \citenamefont {{Luo}},
  \citenamefont {{Zhang}}, \citenamefont {{Li}}, \citenamefont {{Yin}},
  \citenamefont {{Li}}, \citenamefont {{Li}}, \citenamefont {{Dai}},
  \citenamefont {{Li}}, \citenamefont {{Zhang}}, \citenamefont {{Liu}},\ and\
  \citenamefont {{Pan}}}]{2024ApJ...974L..23W}%
  \BibitemOpen
  \bibfield  {author} {\bibinfo {author} {\bibfnamefont {Y.}~\bibnamefont
  {{Wu}}}, \bibinfo {author} {\bibfnamefont {Z.}~\bibnamefont {{Pan}}},
  \bibinfo {author} {\bibfnamefont {L.}~\bibnamefont {{Qian}}}, \bibinfo
  {author} {\bibfnamefont {S.~M.}\ \bibnamefont {{Ransom}}}, \bibinfo {author}
  {\bibfnamefont {R.~P.}\ \bibnamefont {{Eatough}}}, \bibinfo {author}
  {\bibfnamefont {B.}~\bibnamefont {{Wang}}}, \bibinfo {author} {\bibfnamefont
  {P.~C.~C.}\ \bibnamefont {{Freire}}}, \bibinfo {author} {\bibfnamefont
  {K.}~\bibnamefont {{Liu}}}, \bibinfo {author} {\bibfnamefont
  {Z.}~\bibnamefont {{Yan}}}, \bibinfo {author} {\bibfnamefont
  {J.}~\bibnamefont {{Luo}}}, \bibinfo {author} {\bibfnamefont
  {L.}~\bibnamefont {{Zhang}}}, \bibinfo {author} {\bibfnamefont
  {M.}~\bibnamefont {{Li}}}, \bibinfo {author} {\bibfnamefont {D.}~\bibnamefont
  {{Yin}}}, \bibinfo {author} {\bibfnamefont {B.}~\bibnamefont {{Li}}},
  \bibinfo {author} {\bibfnamefont {Y.}~\bibnamefont {{Li}}}, \bibinfo {author}
  {\bibfnamefont {Y.}~\bibnamefont {{Dai}}}, \bibinfo {author} {\bibfnamefont
  {Y.}~\bibnamefont {{Li}}}, \bibinfo {author} {\bibfnamefont {X.}~\bibnamefont
  {{Zhang}}}, \bibinfo {author} {\bibfnamefont {T.}~\bibnamefont {{Liu}}},\
  and\ \bibinfo {author} {\bibfnamefont {Y.}~\bibnamefont {{Pan}}},\ }\bibfield
   {title} {\bibinfo {title} {{The Discovery of Three Pulsars in the Globular
  Cluster M15 with FAST}},\ }\href {https://doi.org/10.3847/2041-8213/ad7b9e}
  {\bibfield  {journal} {\bibinfo  {journal} {\apjl}\ }\textbf {\bibinfo
  {volume} {974}},\ \bibinfo {eid} {L23} (\bibinfo {year} {2024})},\ \Eprint
  {https://arxiv.org/abs/2312.06067} {arXiv:2312.06067 [astro-ph.HE]}
  \BibitemShut {NoStop}%
\bibitem [{\citenamefont {{Pan}}\ \emph {et~al.}(2021)\citenamefont {{Pan}},
  \citenamefont {{Qian}}, \citenamefont {{Ma}}, \citenamefont {{Liu}},
  \citenamefont {{Wang}}, \citenamefont {{Luo}}, \citenamefont {{Yan}},
  \citenamefont {{Ransom}}, \citenamefont {{Lorimer}}, \citenamefont {{Li}},\
  and\ \citenamefont {{Jiang}}}]{2021ApJ...915L..28P}%
  \BibitemOpen
  \bibfield  {author} {\bibinfo {author} {\bibfnamefont {Z.}~\bibnamefont
  {{Pan}}}, \bibinfo {author} {\bibfnamefont {L.}~\bibnamefont {{Qian}}},
  \bibinfo {author} {\bibfnamefont {X.}~\bibnamefont {{Ma}}}, \bibinfo {author}
  {\bibfnamefont {K.}~\bibnamefont {{Liu}}}, \bibinfo {author} {\bibfnamefont
  {L.}~\bibnamefont {{Wang}}}, \bibinfo {author} {\bibfnamefont
  {J.}~\bibnamefont {{Luo}}}, \bibinfo {author} {\bibfnamefont
  {Z.}~\bibnamefont {{Yan}}}, \bibinfo {author} {\bibfnamefont
  {S.}~\bibnamefont {{Ransom}}}, \bibinfo {author} {\bibfnamefont
  {D.}~\bibnamefont {{Lorimer}}}, \bibinfo {author} {\bibfnamefont
  {D.}~\bibnamefont {{Li}}},\ and\ \bibinfo {author} {\bibfnamefont
  {P.}~\bibnamefont {{Jiang}}},\ }\bibfield  {title} {\bibinfo {title} {{FAST
  Globular Cluster Pulsar Survey: Twenty-four Pulsars Discovered in 15 Globular
  Clusters}},\ }\href {https://doi.org/10.3847/2041-8213/ac0bbd} {\bibfield
  {journal} {\bibinfo  {journal} {\apjl}\ }\textbf {\bibinfo {volume} {915}},\
  \bibinfo {eid} {L28} (\bibinfo {year} {2021})},\ \Eprint
  {https://arxiv.org/abs/2106.08559} {arXiv:2106.08559 [astro-ph.HE]}
  \BibitemShut {NoStop}%
\bibitem [{\citenamefont {{Gao}}\ \emph {et~al.}(2024)\citenamefont {{Gao}},
  \citenamefont {{Shao}}, \citenamefont {{Wang}}, \citenamefont {{Zhou}},
  \citenamefont {{Li}}, \citenamefont {{Zhang}}, \citenamefont {{Kania}},
  \citenamefont {{Lorimer}},\ and\ \citenamefont {{Li}}}]{2024ApJ...974L...2G}%
  \BibitemOpen
  \bibfield  {author} {\bibinfo {author} {\bibfnamefont {S.-J.}\ \bibnamefont
  {{Gao}}}, \bibinfo {author} {\bibfnamefont {Y.-X.}\ \bibnamefont {{Shao}}},
  \bibinfo {author} {\bibfnamefont {P.}~\bibnamefont {{Wang}}}, \bibinfo
  {author} {\bibfnamefont {P.}~\bibnamefont {{Zhou}}}, \bibinfo {author}
  {\bibfnamefont {X.-D.}\ \bibnamefont {{Li}}}, \bibinfo {author}
  {\bibfnamefont {L.}~\bibnamefont {{Zhang}}}, \bibinfo {author} {\bibfnamefont
  {J.~W.}\ \bibnamefont {{Kania}}}, \bibinfo {author} {\bibfnamefont {D.~R.}\
  \bibnamefont {{Lorimer}}},\ and\ \bibinfo {author} {\bibfnamefont
  {D.}~\bibnamefont {{Li}}},\ }\bibfield  {title} {\bibinfo {title} {{Discovery
  of a Millisecond Pulsar Associated with Terzan 6}},\ }\href
  {https://doi.org/10.3847/2041-8213/ad7a69} {\bibfield  {journal} {\bibinfo
  {journal} {\apjl}\ }\textbf {\bibinfo {volume} {974}},\ \bibinfo {eid} {L2}
  (\bibinfo {year} {2024})},\ \Eprint {https://arxiv.org/abs/2409.10801}
  {arXiv:2409.10801 [astro-ph.HE]} \BibitemShut {NoStop}%
\bibitem [{\citenamefont {{Lian}}\ \emph {et~al.}(2025)\citenamefont {{Lian}},
  \citenamefont {{Pan}}, \citenamefont {{Zhang}}, \citenamefont {{Cao}},
  \citenamefont {{Freire}}, \citenamefont {{Qian}}, \citenamefont {{Eatough}},
  \citenamefont {{Shao}}, \citenamefont {{Ransom}}, \citenamefont {{Lorimer}},
  \citenamefont {{Yin}}, \citenamefont {{Dai}}, \citenamefont {{Liu}},
  \citenamefont {{Wang}}, \citenamefont {{Wang}}, \citenamefont {{Zhang}},
  \citenamefont {{Feng}}, \citenamefont {{Li}}, \citenamefont {{Li}},
  \citenamefont {{Liu}}, \citenamefont {{Li}}, \citenamefont {{Peng}},
  \citenamefont {{Pan}}, \citenamefont {{Wu}}, \citenamefont {{Zhang}},
  \citenamefont {{Zhang}},\ and\ \citenamefont
  {{Jiang}}}]{2025arXiv250607970L}%
  \BibitemOpen
  \bibfield  {author} {\bibinfo {author} {\bibfnamefont {Y.}~\bibnamefont
  {{Lian}}}, \bibinfo {author} {\bibfnamefont {Z.}~\bibnamefont {{Pan}}},
  \bibinfo {author} {\bibfnamefont {H.}~\bibnamefont {{Zhang}}}, \bibinfo
  {author} {\bibfnamefont {S.}~\bibnamefont {{Cao}}}, \bibinfo {author}
  {\bibfnamefont {P.~C.~C.}\ \bibnamefont {{Freire}}}, \bibinfo {author}
  {\bibfnamefont {L.}~\bibnamefont {{Qian}}}, \bibinfo {author} {\bibfnamefont
  {R.~P.}\ \bibnamefont {{Eatough}}}, \bibinfo {author} {\bibfnamefont
  {L.}~\bibnamefont {{Shao}}}, \bibinfo {author} {\bibfnamefont {S.~M.}\
  \bibnamefont {{Ransom}}}, \bibinfo {author} {\bibfnamefont {D.~R.}\
  \bibnamefont {{Lorimer}}}, \bibinfo {author} {\bibfnamefont {D.}~\bibnamefont
  {{Yin}}}, \bibinfo {author} {\bibfnamefont {Y.}~\bibnamefont {{Dai}}},
  \bibinfo {author} {\bibfnamefont {K.}~\bibnamefont {{Liu}}}, \bibinfo
  {author} {\bibfnamefont {L.}~\bibnamefont {{Wang}}}, \bibinfo {author}
  {\bibfnamefont {Y.}~\bibnamefont {{Wang}}}, \bibinfo {author} {\bibfnamefont
  {Z.}~\bibnamefont {{Zhang}}}, \bibinfo {author} {\bibfnamefont
  {Z.}~\bibnamefont {{Feng}}}, \bibinfo {author} {\bibfnamefont
  {B.}~\bibnamefont {{Li}}}, \bibinfo {author} {\bibfnamefont {M.}~\bibnamefont
  {{Li}}}, \bibinfo {author} {\bibfnamefont {T.}~\bibnamefont {{Liu}}},
  \bibinfo {author} {\bibfnamefont {Y.}~\bibnamefont {{Li}}}, \bibinfo {author}
  {\bibfnamefont {B.}~\bibnamefont {{Peng}}}, \bibinfo {author} {\bibfnamefont
  {Y.}~\bibnamefont {{Pan}}}, \bibinfo {author} {\bibfnamefont
  {Y.}~\bibnamefont {{Wu}}}, \bibinfo {author} {\bibfnamefont {L.}~\bibnamefont
  {{Zhang}}}, \bibinfo {author} {\bibfnamefont {X.}~\bibnamefont {{Zhang}}},\
  and\ \bibinfo {author} {\bibfnamefont {P.}~\bibnamefont {{Jiang}}},\
  }\bibfield  {title} {\bibinfo {title} {{The FAST Globular Cluster Pulsar
  Survey (GC FANS)}},\ }\href {https://doi.org/10.48550/arXiv.2506.07970}
  {\bibfield  {journal} {\bibinfo  {journal} {arXiv e-prints}\ ,\ \bibinfo
  {eid} {arXiv:2506.07970}} (\bibinfo {year} {2025})},\ \Eprint
  {https://arxiv.org/abs/2506.07970} {arXiv:2506.07970 [astro-ph.HE]}
  \BibitemShut {NoStop}%
\bibitem [{\citenamefont {{Ridolfi}}\ \emph {et~al.}(2021)\citenamefont
  {{Ridolfi}}, \citenamefont {{Gautam}}, \citenamefont {{Freire}},
  \citenamefont {{Ransom}}, \citenamefont {{Buchner}}, \citenamefont
  {{Possenti}}, \citenamefont {{Venkatraman Krishnan}}, \citenamefont
  {{Bailes}}, \citenamefont {{Kramer}}, \citenamefont {{Stappers}},
  \citenamefont {{Abbate}}, \citenamefont {{Barr}}, \citenamefont {{Burgay}},
  \citenamefont {{Camilo}}, \citenamefont {{Corongiu}}, \citenamefont
  {{Jameson}}, \citenamefont {{Padmanabh}}, \citenamefont {{Vleeschower}},
  \citenamefont {{Champion}}, \citenamefont {{Chen}}, \citenamefont {{Geyer}},
  \citenamefont {{Karastergiou}}, \citenamefont {{Karuppusamy}}, \citenamefont
  {{Parthasarathy}}, \citenamefont {{Reardon}}, \citenamefont {{Serylak}},
  \citenamefont {{Shannon}},\ and\ \citenamefont
  {{Spiewak}}}]{2021MNRAS.504.1407R}%
  \BibitemOpen
  \bibfield  {author} {\bibinfo {author} {\bibfnamefont {A.}~\bibnamefont
  {{Ridolfi}}}, \bibinfo {author} {\bibfnamefont {T.}~\bibnamefont {{Gautam}}},
  \bibinfo {author} {\bibfnamefont {P.~C.~C.}\ \bibnamefont {{Freire}}},
  \bibinfo {author} {\bibfnamefont {S.~M.}\ \bibnamefont {{Ransom}}}, \bibinfo
  {author} {\bibfnamefont {S.~J.}\ \bibnamefont {{Buchner}}}, \bibinfo {author}
  {\bibfnamefont {A.}~\bibnamefont {{Possenti}}}, \bibinfo {author}
  {\bibfnamefont {V.}~\bibnamefont {{Venkatraman Krishnan}}}, \bibinfo {author}
  {\bibfnamefont {M.}~\bibnamefont {{Bailes}}}, \bibinfo {author}
  {\bibfnamefont {M.}~\bibnamefont {{Kramer}}}, \bibinfo {author}
  {\bibfnamefont {B.~W.}\ \bibnamefont {{Stappers}}}, \bibinfo {author}
  {\bibfnamefont {F.}~\bibnamefont {{Abbate}}}, \bibinfo {author}
  {\bibfnamefont {E.~D.}\ \bibnamefont {{Barr}}}, \bibinfo {author}
  {\bibfnamefont {M.}~\bibnamefont {{Burgay}}}, \bibinfo {author}
  {\bibfnamefont {F.}~\bibnamefont {{Camilo}}}, \bibinfo {author}
  {\bibfnamefont {A.}~\bibnamefont {{Corongiu}}}, \bibinfo {author}
  {\bibfnamefont {A.}~\bibnamefont {{Jameson}}}, \bibinfo {author}
  {\bibfnamefont {P.~V.}\ \bibnamefont {{Padmanabh}}}, \bibinfo {author}
  {\bibfnamefont {L.}~\bibnamefont {{Vleeschower}}}, \bibinfo {author}
  {\bibfnamefont {D.~J.}\ \bibnamefont {{Champion}}}, \bibinfo {author}
  {\bibfnamefont {W.}~\bibnamefont {{Chen}}}, \bibinfo {author} {\bibfnamefont
  {M.}~\bibnamefont {{Geyer}}}, \bibinfo {author} {\bibfnamefont
  {A.}~\bibnamefont {{Karastergiou}}}, \bibinfo {author} {\bibfnamefont
  {R.}~\bibnamefont {{Karuppusamy}}}, \bibinfo {author} {\bibfnamefont
  {A.}~\bibnamefont {{Parthasarathy}}}, \bibinfo {author} {\bibfnamefont
  {D.~J.}\ \bibnamefont {{Reardon}}}, \bibinfo {author} {\bibfnamefont
  {M.}~\bibnamefont {{Serylak}}}, \bibinfo {author} {\bibfnamefont {R.~M.}\
  \bibnamefont {{Shannon}}},\ and\ \bibinfo {author} {\bibfnamefont
  {R.}~\bibnamefont {{Spiewak}}},\ }\bibfield  {title} {\bibinfo {title}
  {{Eight new millisecond pulsars from the first MeerKAT globular cluster
  census}},\ }\href {https://doi.org/10.1093/mnras/stab790} {\bibfield
  {journal} {\bibinfo  {journal} {\mnras}\ }\textbf {\bibinfo {volume} {504}},\
  \bibinfo {pages} {1407} (\bibinfo {year} {2021})},\ \Eprint
  {https://arxiv.org/abs/2103.04800} {arXiv:2103.04800 [astro-ph.HE]}
  \BibitemShut {NoStop}%
\bibitem [{\citenamefont {{Das}}\ \emph {et~al.}(2025)\citenamefont {{Das}},
  \citenamefont {{Roy}}, \citenamefont {{Freire}}, \citenamefont {{Ransom}},
  \citenamefont {{Bhattacharyya}}, \citenamefont {{Ad{\'a}mek}}, \citenamefont
  {{Armour}}, \citenamefont {{Kudale}},\ and\ \citenamefont
  {{Muley}}}]{2025arXiv250209154D}%
  \BibitemOpen
  \bibfield  {author} {\bibinfo {author} {\bibfnamefont {J.}~\bibnamefont
  {{Das}}}, \bibinfo {author} {\bibfnamefont {J.}~\bibnamefont {{Roy}}},
  \bibinfo {author} {\bibfnamefont {P.~C.~C.}\ \bibnamefont {{Freire}}},
  \bibinfo {author} {\bibfnamefont {S.~M.}\ \bibnamefont {{Ransom}}}, \bibinfo
  {author} {\bibfnamefont {B.}~\bibnamefont {{Bhattacharyya}}}, \bibinfo
  {author} {\bibfnamefont {K.}~\bibnamefont {{Ad{\'a}mek}}}, \bibinfo {author}
  {\bibfnamefont {W.}~\bibnamefont {{Armour}}}, \bibinfo {author}
  {\bibfnamefont {S.}~\bibnamefont {{Kudale}}},\ and\ \bibinfo {author}
  {\bibfnamefont {M.~V.}\ \bibnamefont {{Muley}}},\ }\bibfield  {title}
  {\bibinfo {title} {{Globular Clusters GMRT Pulsar Search (GCGPS) I: Survey
  description, discovery and timing of the first pulsar in NGC 6093 (M80)}},\
  }\href {https://doi.org/10.48550/arXiv.2502.09154} {\bibfield  {journal}
  {\bibinfo  {journal} {arXiv e-prints}\ ,\ \bibinfo {eid} {arXiv:2502.09154}}
  (\bibinfo {year} {2025})},\ \Eprint {https://arxiv.org/abs/2502.09154}
  {arXiv:2502.09154 [astro-ph.HE]} \BibitemShut {NoStop}%
\bibitem [{\citenamefont {{Dai}}\ \emph {et~al.}(2020)\citenamefont {{Dai}},
  \citenamefont {{Johnston}}, \citenamefont {{Kerr}}, \citenamefont {{Camilo}},
  \citenamefont {{Cameron}}, \citenamefont {{Toomey}},\ and\ \citenamefont
  {{Kumamoto}}}]{dai20}%
  \BibitemOpen
  \bibfield  {author} {\bibinfo {author} {\bibfnamefont {S.}~\bibnamefont
  {{Dai}}}, \bibinfo {author} {\bibfnamefont {S.}~\bibnamefont {{Johnston}}},
  \bibinfo {author} {\bibfnamefont {M.}~\bibnamefont {{Kerr}}}, \bibinfo
  {author} {\bibfnamefont {F.}~\bibnamefont {{Camilo}}}, \bibinfo {author}
  {\bibfnamefont {A.}~\bibnamefont {{Cameron}}}, \bibinfo {author}
  {\bibfnamefont {L.}~\bibnamefont {{Toomey}}},\ and\ \bibinfo {author}
  {\bibfnamefont {H.}~\bibnamefont {{Kumamoto}}},\ }\bibfield  {title}
  {\bibinfo {title} {{Discovery of Millisecond Pulsars in the Globular Cluster
  Omega Centauri}},\ }\href {https://doi.org/10.3847/2041-8213/ab621a}
  {\bibfield  {journal} {\bibinfo  {journal} {\apjl}\ }\textbf {\bibinfo
  {volume} {888}},\ \bibinfo {eid} {L18} (\bibinfo {year} {2020})},\ \Eprint
  {https://arxiv.org/abs/1912.08079} {arXiv:1912.08079 [astro-ph.HE]}
  \BibitemShut {NoStop}%
\bibitem [{\citenamefont {{Dai}}\ \emph {et~al.}(2023)\citenamefont {{Dai}},
  \citenamefont {{Johnston}}, \citenamefont {{Kerr}}, \citenamefont
  {{Berteaud}}, \citenamefont {{Bhattacharyya}}, \citenamefont {{Camilo}},\
  and\ \citenamefont {{Keane}}}]{dai23}%
  \BibitemOpen
  \bibfield  {author} {\bibinfo {author} {\bibfnamefont {S.}~\bibnamefont
  {{Dai}}}, \bibinfo {author} {\bibfnamefont {S.}~\bibnamefont {{Johnston}}},
  \bibinfo {author} {\bibfnamefont {M.}~\bibnamefont {{Kerr}}}, \bibinfo
  {author} {\bibfnamefont {J.}~\bibnamefont {{Berteaud}}}, \bibinfo {author}
  {\bibfnamefont {B.}~\bibnamefont {{Bhattacharyya}}}, \bibinfo {author}
  {\bibfnamefont {F.}~\bibnamefont {{Camilo}}},\ and\ \bibinfo {author}
  {\bibfnamefont {E.}~\bibnamefont {{Keane}}},\ }\bibfield  {title} {\bibinfo
  {title} {{Timing of pulsars in the globular cluster omega centauri}},\ }\href
  {https://doi.org/10.1093/mnras/stad704} {\bibfield  {journal} {\bibinfo
  {journal} {\mnras}\ }\textbf {\bibinfo {volume} {521}},\ \bibinfo {pages}
  {2616} (\bibinfo {year} {2023})},\ \Eprint {https://arxiv.org/abs/2303.02834}
  {arXiv:2303.02834 [astro-ph.HE]} \BibitemShut {NoStop}%
\bibitem [{\citenamefont {{Hessels}}\ \emph {et~al.}(2015)\citenamefont
  {{Hessels}}, \citenamefont {{Possenti}}, \citenamefont {{Bailes}},
  \citenamefont {{Bassa}}, \citenamefont {{Freire}}, \citenamefont {{Lorimer}},
  \citenamefont {{Lynch}}, \citenamefont {{Ransom}},\ and\ \citenamefont
  {{Stairs}}}]{2015aska.confE..47H}%
  \BibitemOpen
  \bibfield  {author} {\bibinfo {author} {\bibfnamefont {J.}~\bibnamefont
  {{Hessels}}}, \bibinfo {author} {\bibfnamefont {A.}~\bibnamefont
  {{Possenti}}}, \bibinfo {author} {\bibfnamefont {M.}~\bibnamefont
  {{Bailes}}}, \bibinfo {author} {\bibfnamefont {C.}~\bibnamefont {{Bassa}}},
  \bibinfo {author} {\bibfnamefont {P.~C.~C.}\ \bibnamefont {{Freire}}},
  \bibinfo {author} {\bibfnamefont {D.~R.}\ \bibnamefont {{Lorimer}}}, \bibinfo
  {author} {\bibfnamefont {R.}~\bibnamefont {{Lynch}}}, \bibinfo {author}
  {\bibfnamefont {S.~M.}\ \bibnamefont {{Ransom}}},\ and\ \bibinfo {author}
  {\bibfnamefont {I.~H.}\ \bibnamefont {{Stairs}}},\ }\bibfield  {title}
  {\bibinfo {title} {{Pulsars in Globular Clusters with the SKA}},\ }in\ \href
  {https://doi.org/10.22323/1.215.0047} {\emph {\bibinfo {booktitle} {Advancing
  Astrophysics with the Square Kilometre Array (AASKA14)}}}\ (\bibinfo {year}
  {2015})\ p.~\bibinfo {pages} {47},\ \Eprint
  {https://arxiv.org/abs/1501.00086} {arXiv:1501.00086 [astro-ph.HE]}
  \BibitemShut {NoStop}%
\bibitem [{\citenamefont {{Askar}}\ \emph {et~al.}(2023)\citenamefont
  {{Askar}}, \citenamefont {{Baldassare}},\ and\ \citenamefont
  {{Mezcua}}}]{2023arXiv231112118A}%
  \BibitemOpen
  \bibfield  {author} {\bibinfo {author} {\bibfnamefont {A.}~\bibnamefont
  {{Askar}}}, \bibinfo {author} {\bibfnamefont {V.~F.}\ \bibnamefont
  {{Baldassare}}},\ and\ \bibinfo {author} {\bibfnamefont {M.}~\bibnamefont
  {{Mezcua}}},\ }\bibfield  {title} {\bibinfo {title} {{Intermediate-Mass Black
  Holes in Star Clusters and Dwarf Galaxies}},\ }\href
  {https://doi.org/10.48550/arXiv.2311.12118} {\bibfield  {journal} {\bibinfo
  {journal} {arXiv e-prints}\ ,\ \bibinfo {eid} {arXiv:2311.12118}} (\bibinfo
  {year} {2023})},\ \Eprint {https://arxiv.org/abs/2311.12118}
  {arXiv:2311.12118 [astro-ph.GA]} \BibitemShut {NoStop}%
\bibitem [{\citenamefont {{Mezcua}}(2017)}]{2017IJMPD..2630021M}%
  \BibitemOpen
  \bibfield  {author} {\bibinfo {author} {\bibfnamefont {M.}~\bibnamefont
  {{Mezcua}}},\ }\bibfield  {title} {\bibinfo {title} {{Observational evidence
  for intermediate-mass black holes}},\ }\href
  {https://doi.org/10.1142/S021827181730021X} {\bibfield  {journal} {\bibinfo
  {journal} {International Journal of Modern Physics D}\ }\textbf {\bibinfo
  {volume} {26}},\ \bibinfo {eid} {1730021} (\bibinfo {year} {2017})},\ \Eprint
  {https://arxiv.org/abs/1705.09667} {arXiv:1705.09667 [astro-ph.GA]}
  \BibitemShut {NoStop}%
\bibitem [{\citenamefont {{Greene}}\ \emph {et~al.}(2020)\citenamefont
  {{Greene}}, \citenamefont {{Strader}},\ and\ \citenamefont
  {{Ho}}}]{2020ARA&A..58..257G}%
  \BibitemOpen
  \bibfield  {author} {\bibinfo {author} {\bibfnamefont {J.~E.}\ \bibnamefont
  {{Greene}}}, \bibinfo {author} {\bibfnamefont {J.}~\bibnamefont
  {{Strader}}},\ and\ \bibinfo {author} {\bibfnamefont {L.~C.}\ \bibnamefont
  {{Ho}}},\ }\bibfield  {title} {\bibinfo {title} {{Intermediate-Mass Black
  Holes}},\ }\href {https://doi.org/10.1146/annurev-astro-032620-021835}
  {\bibfield  {journal} {\bibinfo  {journal} {\araa}\ }\textbf {\bibinfo
  {volume} {58}},\ \bibinfo {pages} {257} (\bibinfo {year} {2020})},\ \Eprint
  {https://arxiv.org/abs/1911.09678} {arXiv:1911.09678 [astro-ph.GA]}
  \BibitemShut {NoStop}%
\bibitem [{\citenamefont {{Baumgardt}}\ \emph {et~al.}(2004)\citenamefont
  {{Baumgardt}}, \citenamefont {{Makino}},\ and\ \citenamefont
  {{Ebisuzaki}}}]{2004ApJ...613.1143B}%
  \BibitemOpen
  \bibfield  {author} {\bibinfo {author} {\bibfnamefont {H.}~\bibnamefont
  {{Baumgardt}}}, \bibinfo {author} {\bibfnamefont {J.}~\bibnamefont
  {{Makino}}},\ and\ \bibinfo {author} {\bibfnamefont {T.}~\bibnamefont
  {{Ebisuzaki}}},\ }\bibfield  {title} {\bibinfo {title} {{Massive Black Holes
  in Star Clusters. II. Realistic Cluster Models}},\ }\href
  {https://doi.org/10.1086/423299} {\bibfield  {journal} {\bibinfo  {journal}
  {\apj}\ }\textbf {\bibinfo {volume} {613}},\ \bibinfo {pages} {1143}
  (\bibinfo {year} {2004})},\ \Eprint {https://arxiv.org/abs/astro-ph/0406231}
  {arXiv:astro-ph/0406231 [astro-ph]} \BibitemShut {NoStop}%
\bibitem [{\citenamefont {{Lee}}\ \emph {et~al.}(2025)\citenamefont {{Lee}},
  \citenamefont {{Lee}}, \citenamefont {{Kim}}, \citenamefont {{Spurzem}},
  \citenamefont {{Hong}},\ and\ \citenamefont {{Chung}}}]{2025arXiv250322109L}%
  \BibitemOpen
  \bibfield  {author} {\bibinfo {author} {\bibfnamefont {S.}~\bibnamefont
  {{Lee}}}, \bibinfo {author} {\bibfnamefont {H.~M.}\ \bibnamefont {{Lee}}},
  \bibinfo {author} {\bibfnamefont {J.-h.}\ \bibnamefont {{Kim}}}, \bibinfo
  {author} {\bibfnamefont {R.}~\bibnamefont {{Spurzem}}}, \bibinfo {author}
  {\bibfnamefont {J.}~\bibnamefont {{Hong}}},\ and\ \bibinfo {author}
  {\bibfnamefont {E.}~\bibnamefont {{Chung}}},\ }\bibfield  {title} {\bibinfo
  {title} {{Formation and Evolution of Compact Binaries Containing Intermediate
  Mass Black Holes in Dense Star Clusters`}},\ }\href
  {https://doi.org/10.48550/arXiv.2503.22109} {\bibfield  {journal} {\bibinfo
  {journal} {arXiv e-prints}\ ,\ \bibinfo {eid} {arXiv:2503.22109}} (\bibinfo
  {year} {2025})},\ \Eprint {https://arxiv.org/abs/2503.22109}
  {arXiv:2503.22109 [astro-ph.GA]} \BibitemShut {NoStop}%
\bibitem [{\citenamefont {{Giersz}}\ \emph {et~al.}(2015)\citenamefont
  {{Giersz}}, \citenamefont {{Leigh}}, \citenamefont {{Hypki}}, \citenamefont
  {{L{\"u}tzgendorf}},\ and\ \citenamefont {{Askar}}}]{2015MNRAS.454.3150G}%
  \BibitemOpen
  \bibfield  {author} {\bibinfo {author} {\bibfnamefont {M.}~\bibnamefont
  {{Giersz}}}, \bibinfo {author} {\bibfnamefont {N.}~\bibnamefont {{Leigh}}},
  \bibinfo {author} {\bibfnamefont {A.}~\bibnamefont {{Hypki}}}, \bibinfo
  {author} {\bibfnamefont {N.}~\bibnamefont {{L{\"u}tzgendorf}}},\ and\
  \bibinfo {author} {\bibfnamefont {A.}~\bibnamefont {{Askar}}},\ }\bibfield
  {title} {\bibinfo {title} {{MOCCA code for star cluster simulations - IV. A
  new scenario for intermediate mass black hole formation in globular
  clusters}},\ }\href {https://doi.org/10.1093/mnras/stv2162} {\bibfield
  {journal} {\bibinfo  {journal} {\mnras}\ }\textbf {\bibinfo {volume} {454}},\
  \bibinfo {pages} {3150} (\bibinfo {year} {2015})},\ \Eprint
  {https://arxiv.org/abs/1506.05234} {arXiv:1506.05234 [astro-ph.GA]}
  \BibitemShut {NoStop}%
\bibitem [{\citenamefont {{Blecha}}\ \emph {et~al.}(2006)\citenamefont
  {{Blecha}}, \citenamefont {{Ivanova}}, \citenamefont {{Kalogera}},
  \citenamefont {{Belczynski}}, \citenamefont {{Fregeau}},\ and\ \citenamefont
  {{Rasio}}}]{2006ApJ...642..427B}%
  \BibitemOpen
  \bibfield  {author} {\bibinfo {author} {\bibfnamefont {L.}~\bibnamefont
  {{Blecha}}}, \bibinfo {author} {\bibfnamefont {N.}~\bibnamefont {{Ivanova}}},
  \bibinfo {author} {\bibfnamefont {V.}~\bibnamefont {{Kalogera}}}, \bibinfo
  {author} {\bibfnamefont {K.}~\bibnamefont {{Belczynski}}}, \bibinfo {author}
  {\bibfnamefont {J.}~\bibnamefont {{Fregeau}}},\ and\ \bibinfo {author}
  {\bibfnamefont {F.}~\bibnamefont {{Rasio}}},\ }\bibfield  {title} {\bibinfo
  {title} {{Close Binary Interactions of Intermediate-Mass Black Holes:
  Possible Ultraluminous X-Ray Sources?}},\ }\href
  {https://doi.org/10.1086/500727} {\bibfield  {journal} {\bibinfo  {journal}
  {\apj}\ }\textbf {\bibinfo {volume} {642}},\ \bibinfo {pages} {427} (\bibinfo
  {year} {2006})},\ \Eprint {https://arxiv.org/abs/astro-ph/0508597}
  {arXiv:astro-ph/0508597 [astro-ph]} \BibitemShut {NoStop}%
\bibitem [{\citenamefont {{Leigh}}\ \emph {et~al.}(2014)\citenamefont
  {{Leigh}}, \citenamefont {{L{\"u}tzgendorf}}, \citenamefont {{Geller}},
  \citenamefont {{Maccarone}}, \citenamefont {{Heinke}},\ and\ \citenamefont
  {{Sesana}}}]{2014MNRAS.444...29L}%
  \BibitemOpen
  \bibfield  {author} {\bibinfo {author} {\bibfnamefont {N.~W.~C.}\
  \bibnamefont {{Leigh}}}, \bibinfo {author} {\bibfnamefont {N.}~\bibnamefont
  {{L{\"u}tzgendorf}}}, \bibinfo {author} {\bibfnamefont {A.~M.}\ \bibnamefont
  {{Geller}}}, \bibinfo {author} {\bibfnamefont {T.~J.}\ \bibnamefont
  {{Maccarone}}}, \bibinfo {author} {\bibfnamefont {C.}~\bibnamefont
  {{Heinke}}},\ and\ \bibinfo {author} {\bibfnamefont {A.}~\bibnamefont
  {{Sesana}}},\ }\bibfield  {title} {\bibinfo {title} {{On the coexistence of
  stellar-mass and intermediate-mass black holes in globular clusters}},\
  }\href {https://doi.org/10.1093/mnras/stu1437} {\bibfield  {journal}
  {\bibinfo  {journal} {\mnras}\ }\textbf {\bibinfo {volume} {444}},\ \bibinfo
  {pages} {29} (\bibinfo {year} {2014})},\ \Eprint
  {https://arxiv.org/abs/1407.4459} {arXiv:1407.4459 [astro-ph.SR]}
  \BibitemShut {NoStop}%
\bibitem [{\citenamefont {{MacLeod}}\ \emph {et~al.}(2016)\citenamefont
  {{MacLeod}}, \citenamefont {{Trenti}},\ and\ \citenamefont
  {{Ramirez-Ruiz}}}]{2016ApJ...819...70M}%
  \BibitemOpen
  \bibfield  {author} {\bibinfo {author} {\bibfnamefont {M.}~\bibnamefont
  {{MacLeod}}}, \bibinfo {author} {\bibfnamefont {M.}~\bibnamefont
  {{Trenti}}},\ and\ \bibinfo {author} {\bibfnamefont {E.}~\bibnamefont
  {{Ramirez-Ruiz}}},\ }\bibfield  {title} {\bibinfo {title} {{The Close Stellar
  Companions to Intermediate-mass Black Holes}},\ }\href
  {https://doi.org/10.3847/0004-637X/819/1/70} {\bibfield  {journal} {\bibinfo
  {journal} {\apj}\ }\textbf {\bibinfo {volume} {819}},\ \bibinfo {eid} {70}
  (\bibinfo {year} {2016})},\ \Eprint {https://arxiv.org/abs/1508.07000}
  {arXiv:1508.07000 [astro-ph.HE]} \BibitemShut {NoStop}%
\bibitem [{\citenamefont {{Arca Sedda}}\ \emph {et~al.}(2021)\citenamefont
  {{Arca Sedda}}, \citenamefont {{Amaro Seoane}},\ and\ \citenamefont
  {{Chen}}}]{2021A&A...652A..54A}%
  \BibitemOpen
  \bibfield  {author} {\bibinfo {author} {\bibfnamefont {M.}~\bibnamefont
  {{Arca Sedda}}}, \bibinfo {author} {\bibfnamefont {P.}~\bibnamefont {{Amaro
  Seoane}}},\ and\ \bibinfo {author} {\bibfnamefont {X.}~\bibnamefont
  {{Chen}}},\ }\bibfield  {title} {\bibinfo {title} {{Merging stellar and
  intermediate-mass black holes in dense clusters: implications for LIGO, LISA,
  and the next generation of gravitational wave detectors}},\ }\href
  {https://doi.org/10.1051/0004-6361/202037785} {\bibfield  {journal} {\bibinfo
   {journal} {\aap}\ }\textbf {\bibinfo {volume} {652}},\ \bibinfo {eid} {A54}
  (\bibinfo {year} {2021})},\ \Eprint {https://arxiv.org/abs/2007.13746}
  {arXiv:2007.13746 [astro-ph.GA]} \BibitemShut {NoStop}%
\bibitem [{\citenamefont {{Konstantinidis}}\ \emph {et~al.}(2013)\citenamefont
  {{Konstantinidis}}, \citenamefont {{Amaro-Seoane}},\ and\ \citenamefont
  {{Kokkotas}}}]{2013A&A...557A.135K}%
  \BibitemOpen
  \bibfield  {author} {\bibinfo {author} {\bibfnamefont {S.}~\bibnamefont
  {{Konstantinidis}}}, \bibinfo {author} {\bibfnamefont {P.}~\bibnamefont
  {{Amaro-Seoane}}},\ and\ \bibinfo {author} {\bibfnamefont {K.~D.}\
  \bibnamefont {{Kokkotas}}},\ }\bibfield  {title} {\bibinfo {title}
  {{Investigating the retention of intermediate-mass black holes in star
  clusters using N-body simulations}},\ }\href
  {https://doi.org/10.1051/0004-6361/201219620} {\bibfield  {journal} {\bibinfo
   {journal} {\aap}\ }\textbf {\bibinfo {volume} {557}},\ \bibinfo {eid} {A135}
  (\bibinfo {year} {2013})},\ \Eprint {https://arxiv.org/abs/1108.5175}
  {arXiv:1108.5175 [astro-ph.CO]} \BibitemShut {NoStop}%
\bibitem [{\citenamefont {{Haster}}\ \emph {et~al.}(2016)\citenamefont
  {{Haster}}, \citenamefont {{Antonini}}, \citenamefont {{Kalogera}},\ and\
  \citenamefont {{Mandel}}}]{2016ApJ...832..192H}%
  \BibitemOpen
  \bibfield  {author} {\bibinfo {author} {\bibfnamefont {C.-J.}\ \bibnamefont
  {{Haster}}}, \bibinfo {author} {\bibfnamefont {F.}~\bibnamefont
  {{Antonini}}}, \bibinfo {author} {\bibfnamefont {V.}~\bibnamefont
  {{Kalogera}}},\ and\ \bibinfo {author} {\bibfnamefont {I.}~\bibnamefont
  {{Mandel}}},\ }\bibfield  {title} {\bibinfo {title} {{N-Body Dynamics of
  Intermediate Mass-ratio Inspirals in Star Clusters}},\ }\href
  {https://doi.org/10.3847/0004-637X/832/2/192} {\bibfield  {journal} {\bibinfo
   {journal} {\apj}\ }\textbf {\bibinfo {volume} {832}},\ \bibinfo {eid} {192}
  (\bibinfo {year} {2016})},\ \Eprint {https://arxiv.org/abs/1606.07097}
  {arXiv:1606.07097 [astro-ph.HE]} \BibitemShut {NoStop}%
\bibitem [{\citenamefont {{G{\"u}rkan}}\ \emph {et~al.}(2006)\citenamefont
  {{G{\"u}rkan}}, \citenamefont {{Fregeau}},\ and\ \citenamefont
  {{Rasio}}}]{2006ApJ...640L..39G}%
  \BibitemOpen
  \bibfield  {author} {\bibinfo {author} {\bibfnamefont {M.~A.}\ \bibnamefont
  {{G{\"u}rkan}}}, \bibinfo {author} {\bibfnamefont {J.~M.}\ \bibnamefont
  {{Fregeau}}},\ and\ \bibinfo {author} {\bibfnamefont {F.~A.}\ \bibnamefont
  {{Rasio}}},\ }\bibfield  {title} {\bibinfo {title} {{Massive Black Hole
  Binaries from Collisional Runaways}},\ }\href
  {https://doi.org/10.1086/503295} {\bibfield  {journal} {\bibinfo  {journal}
  {\apjl}\ }\textbf {\bibinfo {volume} {640}},\ \bibinfo {pages} {L39}
  (\bibinfo {year} {2006})},\ \Eprint {https://arxiv.org/abs/astro-ph/0512642}
  {arXiv:astro-ph/0512642 [astro-ph]} \BibitemShut {NoStop}%
\bibitem [{\citenamefont {{Amaro-Seoane}}\ and\ \citenamefont
  {{Freitag}}(2006)}]{2006ApJ...653L..53A}%
  \BibitemOpen
  \bibfield  {author} {\bibinfo {author} {\bibfnamefont {P.}~\bibnamefont
  {{Amaro-Seoane}}}\ and\ \bibinfo {author} {\bibfnamefont {M.}~\bibnamefont
  {{Freitag}}},\ }\bibfield  {title} {\bibinfo {title} {{Intermediate-Mass
  Black Holes in Colliding Clusters: Implications for Lower Frequency
  Gravitational-Wave Astronomy}},\ }\href {https://doi.org/10.1086/510405}
  {\bibfield  {journal} {\bibinfo  {journal} {\apjl}\ }\textbf {\bibinfo
  {volume} {653}},\ \bibinfo {pages} {L53} (\bibinfo {year} {2006})},\ \Eprint
  {https://arxiv.org/abs/astro-ph/0610478} {arXiv:astro-ph/0610478 [astro-ph]}
  \BibitemShut {NoStop}%
\bibitem [{\citenamefont {{Arca Sedda}}\ and\ \citenamefont
  {{Mastrobuono-Battisti}}(2019)}]{2019arXiv190605864A}%
  \BibitemOpen
  \bibfield  {author} {\bibinfo {author} {\bibfnamefont {M.}~\bibnamefont
  {{Arca Sedda}}}\ and\ \bibinfo {author} {\bibfnamefont {A.}~\bibnamefont
  {{Mastrobuono-Battisti}}},\ }\bibfield  {title} {\bibinfo {title} {{Mergers
  of globular clusters in the Galactic disc: intermediate mass black hole
  coalescence and implications for gravitational waves detection}},\ }\href
  {https://doi.org/10.48550/arXiv.1906.05864} {\bibfield  {journal} {\bibinfo
  {journal} {arXiv e-prints}\ ,\ \bibinfo {eid} {arXiv:1906.05864}} (\bibinfo
  {year} {2019})},\ \Eprint {https://arxiv.org/abs/1906.05864}
  {arXiv:1906.05864 [astro-ph.GA]} \BibitemShut {NoStop}%
\bibitem [{\citenamefont {{Rasskazov}}\ \emph {et~al.}(2020)\citenamefont
  {{Rasskazov}}, \citenamefont {{Fragione}},\ and\ \citenamefont
  {{Kocsis}}}]{2020ApJ...899..149R}%
  \BibitemOpen
  \bibfield  {author} {\bibinfo {author} {\bibfnamefont {A.}~\bibnamefont
  {{Rasskazov}}}, \bibinfo {author} {\bibfnamefont {G.}~\bibnamefont
  {{Fragione}}},\ and\ \bibinfo {author} {\bibfnamefont {B.}~\bibnamefont
  {{Kocsis}}},\ }\bibfield  {title} {\bibinfo {title} {{Binary
  Intermediate-mass Black Hole Mergers in Globular Clusters}},\ }\href
  {https://doi.org/10.3847/1538-4357/aba2f4} {\bibfield  {journal} {\bibinfo
  {journal} {\apj}\ }\textbf {\bibinfo {volume} {899}},\ \bibinfo {eid} {149}
  (\bibinfo {year} {2020})},\ \Eprint {https://arxiv.org/abs/1912.07681}
  {arXiv:1912.07681 [astro-ph.GA]} \BibitemShut {NoStop}%
\bibitem [{\citenamefont {{Souvaitzis}}\ \emph {et~al.}(2025)\citenamefont
  {{Souvaitzis}}, \citenamefont {{Rantala}},\ and\ \citenamefont
  {{Naab}}}]{2025MNRAS.tmp..434S}%
  \BibitemOpen
  \bibfield  {author} {\bibinfo {author} {\bibfnamefont {L.}~\bibnamefont
  {{Souvaitzis}}}, \bibinfo {author} {\bibfnamefont {A.}~\bibnamefont
  {{Rantala}}},\ and\ \bibinfo {author} {\bibfnamefont {T.}~\bibnamefont
  {{Naab}}},\ }\bibfield  {title} {\bibinfo {title} {{The role of Massive Black
  Holes in merging star clusters: dynamical evolution, stellar \& compact
  object ejections and gravitational waves}},\ }\bibfield  {journal} {\bibinfo
  {journal} {\mnras}\ }\href {https://doi.org/10.1093/mnras/staf458}
  {10.1093/mnras/staf458} (\bibinfo {year} {2025}),\ \Eprint
  {https://arxiv.org/abs/2503.11813} {arXiv:2503.11813 [astro-ph.GA]}
  \BibitemShut {NoStop}%
\bibitem [{\citenamefont {{Sazhin}}\ and\ \citenamefont
  {{Saphonova}}(1993)}]{1993Ap&SS.208...93S}%
  \BibitemOpen
  \bibfield  {author} {\bibinfo {author} {\bibfnamefont {M.~V.}\ \bibnamefont
  {{Sazhin}}}\ and\ \bibinfo {author} {\bibfnamefont {M.~V.}\ \bibnamefont
  {{Saphonova}}},\ }\bibfield  {title} {\bibinfo {title} {{Gravitational Action
  of Binaries on Pulsar Timing in Globular Clusters}},\ }\href
  {https://doi.org/10.1007/BF00658136} {\bibfield  {journal} {\bibinfo
  {journal} {\apss}\ }\textbf {\bibinfo {volume} {208}},\ \bibinfo {pages} {93}
  (\bibinfo {year} {1993})}\BibitemShut {NoStop}%
\bibitem [{\citenamefont {{Fakir}}(1994)}]{1994PhRvD..50.3795F}%
  \BibitemOpen
  \bibfield  {author} {\bibinfo {author} {\bibfnamefont {R.}~\bibnamefont
  {{Fakir}}},\ }\bibfield  {title} {\bibinfo {title} {{Detectable time delays
  from gravity waves\textbackslash?}},\ }\href
  {https://doi.org/10.1103/PhysRevD.50.3795} {\bibfield  {journal} {\bibinfo
  {journal} {\prd}\ }\textbf {\bibinfo {volume} {50}},\ \bibinfo {pages} {3795}
  (\bibinfo {year} {1994})},\ \Eprint {https://arxiv.org/abs/astro-ph/9306019}
  {arXiv:astro-ph/9306019 [astro-ph]} \BibitemShut {NoStop}%
\bibitem [{\citenamefont {{Lommen}}\ \emph {et~al.}(2005)\citenamefont
  {{Lommen}}, \citenamefont {{Bilikova}}, \citenamefont {{Jenet}},
  \citenamefont {{Portegies Zwart}},\ and\ \citenamefont
  {{Stappers}}}]{lommen05}%
  \BibitemOpen
  \bibfield  {author} {\bibinfo {author} {\bibfnamefont {A.~N.}\ \bibnamefont
  {{Lommen}}}, \bibinfo {author} {\bibfnamefont {J.}~\bibnamefont
  {{Bilikova}}}, \bibinfo {author} {\bibfnamefont {F.~A.}\ \bibnamefont
  {{Jenet}}}, \bibinfo {author} {\bibfnamefont {S.}~\bibnamefont {{Portegies
  Zwart}}},\ and\ \bibinfo {author} {\bibfnamefont {B.~W.}\ \bibnamefont
  {{Stappers}}},\ }\bibfield  {title} {\bibinfo {title} {{Using Pulsars to
  Detect Black Hole Binaries in Globular Clusters}},\ }in\ \href@noop {} {\emph
  {\bibinfo {booktitle} {Binary Radio Pulsars}}},\ \bibinfo {series}
  {Astronomical Society of the Pacific Conference Series}, Vol.\ \bibinfo
  {volume} {328},\ \bibinfo {editor} {edited by\ \bibinfo {editor}
  {\bibfnamefont {F.~A.}\ \bibnamefont {{Rasio}}}\ and\ \bibinfo {editor}
  {\bibfnamefont {I.~H.}\ \bibnamefont {{Stairs}}}}\ (\bibinfo {year} {2005})\
  p.\ \bibinfo {pages} {225}\BibitemShut {NoStop}%
\bibitem [{\citenamefont {{Damour}}\ and\ \citenamefont
  {{Esposito-Far{\`e}se}}(1998)}]{1998PhRvD..58d4003D}%
  \BibitemOpen
  \bibfield  {author} {\bibinfo {author} {\bibfnamefont {T.}~\bibnamefont
  {{Damour}}}\ and\ \bibinfo {author} {\bibfnamefont {G.}~\bibnamefont
  {{Esposito-Far{\`e}se}}},\ }\bibfield  {title} {\bibinfo {title} {{Light
  deflection by gravitational waves from localized sources}},\ }\href
  {https://doi.org/10.1103/PhysRevD.58.044003} {\bibfield  {journal} {\bibinfo
  {journal} {\prd}\ }\textbf {\bibinfo {volume} {58}},\ \bibinfo {eid} {044003}
  (\bibinfo {year} {1998})},\ \Eprint {https://arxiv.org/abs/gr-qc/9802019}
  {arXiv:gr-qc/9802019 [gr-qc]} \BibitemShut {NoStop}%
\bibitem [{\citenamefont {{Kopeikin}}\ \emph {et~al.}(1999)\citenamefont
  {{Kopeikin}}, \citenamefont {{Sch{\"a}fer}}, \citenamefont {{Gwinn}},\ and\
  \citenamefont {{Eubanks}}}]{1999PhRvD..59h4023K}%
  \BibitemOpen
  \bibfield  {author} {\bibinfo {author} {\bibfnamefont {S.~M.}\ \bibnamefont
  {{Kopeikin}}}, \bibinfo {author} {\bibfnamefont {G.}~\bibnamefont
  {{Sch{\"a}fer}}}, \bibinfo {author} {\bibfnamefont {C.~R.}\ \bibnamefont
  {{Gwinn}}},\ and\ \bibinfo {author} {\bibfnamefont {T.~M.}\ \bibnamefont
  {{Eubanks}}},\ }\bibfield  {title} {\bibinfo {title} {{Astrometric and timing
  effects of gravitational waves from localized sources}},\ }\href
  {https://doi.org/10.1103/PhysRevD.59.084023} {\bibfield  {journal} {\bibinfo
  {journal} {\prd}\ }\textbf {\bibinfo {volume} {59}},\ \bibinfo {eid} {084023}
  (\bibinfo {year} {1999})},\ \Eprint {https://arxiv.org/abs/gr-qc/9811003}
  {arXiv:gr-qc/9811003 [gr-qc]} \BibitemShut {NoStop}%
\bibitem [{\citenamefont {{Helfer}}(2013)}]{2013MNRAS.430..305H}%
  \BibitemOpen
  \bibfield  {author} {\bibinfo {author} {\bibfnamefont {A.~D.}\ \bibnamefont
  {{Helfer}}},\ }\bibfield  {title} {\bibinfo {title} {{Light rays,
  gravitational waves and pulse-time offsets}},\ }\href
  {https://doi.org/10.1093/mnras/sts618} {\bibfield  {journal} {\bibinfo
  {journal} {\mnras}\ }\textbf {\bibinfo {volume} {430}},\ \bibinfo {pages}
  {305} (\bibinfo {year} {2013})},\ \Eprint {https://arxiv.org/abs/1212.2926}
  {arXiv:1212.2926 [gr-qc]} \BibitemShut {NoStop}%
\bibitem [{\citenamefont {{Jenet}}\ \emph {et~al.}(2005)\citenamefont
  {{Jenet}}, \citenamefont {{Creighton}},\ and\ \citenamefont
  {{Lommen}}}]{jenet05}%
  \BibitemOpen
  \bibfield  {author} {\bibinfo {author} {\bibfnamefont {F.~A.}\ \bibnamefont
  {{Jenet}}}, \bibinfo {author} {\bibfnamefont {T.}~\bibnamefont
  {{Creighton}}},\ and\ \bibinfo {author} {\bibfnamefont {A.}~\bibnamefont
  {{Lommen}}},\ }\bibfield  {title} {\bibinfo {title} {{Pulsar Timing and the
  Detection of Black Hole Binary Systems in Globular Clusters}},\ }\href
  {https://doi.org/10.1086/431949} {\bibfield  {journal} {\bibinfo  {journal}
  {\apjl}\ }\textbf {\bibinfo {volume} {627}},\ \bibinfo {pages} {L125}
  (\bibinfo {year} {2005})},\ \Eprint {https://arxiv.org/abs/astro-ph/0505585}
  {arXiv:astro-ph/0505585 [astro-ph]} \BibitemShut {NoStop}%
\bibitem [{\citenamefont {{Madison}}\ \emph {et~al.}(2017)\citenamefont
  {{Madison}}, \citenamefont {{Chernoff}},\ and\ \citenamefont
  {{Cordes}}}]{2017PhRvD..96l3016M}%
  \BibitemOpen
  \bibfield  {author} {\bibinfo {author} {\bibfnamefont {D.~R.}\ \bibnamefont
  {{Madison}}}, \bibinfo {author} {\bibfnamefont {D.~F.}\ \bibnamefont
  {{Chernoff}}},\ and\ \bibinfo {author} {\bibfnamefont {J.~M.}\ \bibnamefont
  {{Cordes}}},\ }\bibfield  {title} {\bibinfo {title} {{Pulsar timing
  perturbations from Galactic gravitational wave bursts with memory}},\ }\href
  {https://doi.org/10.1103/PhysRevD.96.123016} {\bibfield  {journal} {\bibinfo
  {journal} {\prd}\ }\textbf {\bibinfo {volume} {96}},\ \bibinfo {eid} {123016}
  (\bibinfo {year} {2017})},\ \Eprint {https://arxiv.org/abs/1710.04974}
  {arXiv:1710.04974 [astro-ph.GA]} \BibitemShut {NoStop}%
\bibitem [{\citenamefont {{Perera}}\ \emph {et~al.}(2017)\citenamefont
  {{Perera}}, \citenamefont {{Stappers}}, \citenamefont {{Lyne}}, \citenamefont
  {{Bassa}}, \citenamefont {{Cognard}}, \citenamefont {{Guillemot}},
  \citenamefont {{Kramer}}, \citenamefont {{Theureau}},\ and\ \citenamefont
  {{Desvignes}}}]{2017MNRAS.468.2114P}%
  \BibitemOpen
  \bibfield  {author} {\bibinfo {author} {\bibfnamefont {B.~B.~P.}\
  \bibnamefont {{Perera}}}, \bibinfo {author} {\bibfnamefont {B.~W.}\
  \bibnamefont {{Stappers}}}, \bibinfo {author} {\bibfnamefont {A.~G.}\
  \bibnamefont {{Lyne}}}, \bibinfo {author} {\bibfnamefont {C.~G.}\
  \bibnamefont {{Bassa}}}, \bibinfo {author} {\bibfnamefont {I.}~\bibnamefont
  {{Cognard}}}, \bibinfo {author} {\bibfnamefont {L.}~\bibnamefont
  {{Guillemot}}}, \bibinfo {author} {\bibfnamefont {M.}~\bibnamefont
  {{Kramer}}}, \bibinfo {author} {\bibfnamefont {G.}~\bibnamefont
  {{Theureau}}},\ and\ \bibinfo {author} {\bibfnamefont {G.}~\bibnamefont
  {{Desvignes}}},\ }\bibfield  {title} {\bibinfo {title} {{Evidence for an
  intermediate-mass black hole in the globular cluster NGC 6624}},\ }\href
  {https://doi.org/10.1093/mnras/stx501} {\bibfield  {journal} {\bibinfo
  {journal} {\mnras}\ }\textbf {\bibinfo {volume} {468}},\ \bibinfo {pages}
  {2114} (\bibinfo {year} {2017})},\ \Eprint {https://arxiv.org/abs/1705.01612}
  {arXiv:1705.01612 [astro-ph.HE]} \BibitemShut {NoStop}%
\bibitem [{\citenamefont {Häberle}\ \emph {et~al.}(2024)\citenamefont
  {Häberle}, \citenamefont {Neumayer}, \citenamefont {Seth}, \citenamefont
  {Bellini}, \citenamefont {Libralato}, \citenamefont {Baumgardt},
  \citenamefont {Whitaker}, \citenamefont {Dumont}, \citenamefont
  {Alfaro-Cuello}, \citenamefont {Anderson}, \citenamefont {Clontz},
  \citenamefont {Kacharov}, \citenamefont {Kamann}, \citenamefont
  {Feldmeier-Krause}, \citenamefont {Milone}, \citenamefont {Nitschai},
  \citenamefont {Pechetti},\ and\ \citenamefont {van~de Ven}}]{Haberle_2024}%
  \BibitemOpen
  \bibfield  {author} {\bibinfo {author} {\bibfnamefont {M.}~\bibnamefont
  {Häberle}}, \bibinfo {author} {\bibfnamefont {N.}~\bibnamefont {Neumayer}},
  \bibinfo {author} {\bibfnamefont {A.}~\bibnamefont {Seth}}, \bibinfo {author}
  {\bibfnamefont {A.}~\bibnamefont {Bellini}}, \bibinfo {author} {\bibfnamefont
  {M.}~\bibnamefont {Libralato}}, \bibinfo {author} {\bibfnamefont
  {H.}~\bibnamefont {Baumgardt}}, \bibinfo {author} {\bibfnamefont
  {M.}~\bibnamefont {Whitaker}}, \bibinfo {author} {\bibfnamefont
  {A.}~\bibnamefont {Dumont}}, \bibinfo {author} {\bibfnamefont
  {M.}~\bibnamefont {Alfaro-Cuello}}, \bibinfo {author} {\bibfnamefont
  {J.}~\bibnamefont {Anderson}}, \bibinfo {author} {\bibfnamefont
  {C.}~\bibnamefont {Clontz}}, \bibinfo {author} {\bibfnamefont
  {N.}~\bibnamefont {Kacharov}}, \bibinfo {author} {\bibfnamefont
  {S.}~\bibnamefont {Kamann}}, \bibinfo {author} {\bibfnamefont
  {A.}~\bibnamefont {Feldmeier-Krause}}, \bibinfo {author} {\bibfnamefont
  {A.}~\bibnamefont {Milone}}, \bibinfo {author} {\bibfnamefont {M.~S.}\
  \bibnamefont {Nitschai}}, \bibinfo {author} {\bibfnamefont {R.}~\bibnamefont
  {Pechetti}},\ and\ \bibinfo {author} {\bibfnamefont {G.}~\bibnamefont {van~de
  Ven}},\ }\bibfield  {title} {\bibinfo {title} {Fast-moving stars around an
  intermediate-mass black hole in $\omega$ centauri},\ }\href
  {https://doi.org/10.1038/s41586-024-07511-z} {\bibfield  {journal} {\bibinfo
  {journal} {Nature}\ }\textbf {\bibinfo {volume} {631}},\ \bibinfo {pages}
  {285–288} (\bibinfo {year} {2024})}\BibitemShut {NoStop}%
\bibitem [{\citenamefont {{Huang}}\ \emph {et~al.}(2025)\citenamefont
  {{Huang}}, \citenamefont {{Li}}, \citenamefont {{Liu}}, \citenamefont
  {{Dong}}, \citenamefont {{Zhang}}, \citenamefont {{Lu}},\ and\ \citenamefont
  {{Du}}}]{2025NSRev..12..347H}%
  \BibitemOpen
  \bibfield  {author} {\bibinfo {author} {\bibfnamefont {Y.}~\bibnamefont
  {{Huang}}}, \bibinfo {author} {\bibfnamefont {Q.}~\bibnamefont {{Li}}},
  \bibinfo {author} {\bibfnamefont {J.}~\bibnamefont {{Liu}}}, \bibinfo
  {author} {\bibfnamefont {X.}~\bibnamefont {{Dong}}}, \bibinfo {author}
  {\bibfnamefont {H.}~\bibnamefont {{Zhang}}}, \bibinfo {author} {\bibfnamefont
  {Y.}~\bibnamefont {{Lu}}},\ and\ \bibinfo {author} {\bibfnamefont
  {C.}~\bibnamefont {{Du}}},\ }\bibfield  {title} {\bibinfo {title} {{A
  high-velocity star recently ejected by an intermediate-mass black hole in
  M15}},\ }\href {https://doi.org/10.1093/nsr/nwae347} {\bibfield  {journal}
  {\bibinfo  {journal} {National Science Review}\ }\textbf {\bibinfo {volume}
  {12}},\ \bibinfo {eid} {347} (\bibinfo {year} {2025})},\ \Eprint
  {https://arxiv.org/abs/2406.00923} {arXiv:2406.00923 [astro-ph.GA]}
  \BibitemShut {NoStop}%
\bibitem [{\citenamefont {{Anholm}}\ \emph {et~al.}(2009)\citenamefont
  {{Anholm}}, \citenamefont {{Ballmer}}, \citenamefont {{Creighton}},
  \citenamefont {{Price}},\ and\ \citenamefont {{Siemens}}}]{Anholm_2009}%
  \BibitemOpen
  \bibfield  {author} {\bibinfo {author} {\bibfnamefont {M.}~\bibnamefont
  {{Anholm}}}, \bibinfo {author} {\bibfnamefont {S.}~\bibnamefont {{Ballmer}}},
  \bibinfo {author} {\bibfnamefont {J.~D.~E.}\ \bibnamefont {{Creighton}}},
  \bibinfo {author} {\bibfnamefont {L.~R.}\ \bibnamefont {{Price}}},\ and\
  \bibinfo {author} {\bibfnamefont {X.}~\bibnamefont {{Siemens}}},\ }\bibfield
  {title} {\bibinfo {title} {{Optimal strategies for gravitational wave
  stochastic background searches in pulsar timing data}},\ }\href
  {https://doi.org/10.1103/PhysRevD.79.084030} {\bibfield  {journal} {\bibinfo
  {journal} {\prd}\ }\textbf {\bibinfo {volume} {79}},\ \bibinfo {eid} {084030}
  (\bibinfo {year} {2009})},\ \Eprint {https://arxiv.org/abs/0809.0701}
  {arXiv:0809.0701 [gr-qc]} \BibitemShut {NoStop}%
\bibitem [{\citenamefont {{Prager}}\ \emph {et~al.}(2017)\citenamefont
  {{Prager}}, \citenamefont {{Ransom}}, \citenamefont {{Freire}}, \citenamefont
  {{Hessels}}, \citenamefont {{Stairs}}, \citenamefont {{Arras}},\ and\
  \citenamefont {{Cadelano}}}]{2017ApJ...845..148P}%
  \BibitemOpen
  \bibfield  {author} {\bibinfo {author} {\bibfnamefont {B.~J.}\ \bibnamefont
  {{Prager}}}, \bibinfo {author} {\bibfnamefont {S.~M.}\ \bibnamefont
  {{Ransom}}}, \bibinfo {author} {\bibfnamefont {P.~C.~C.}\ \bibnamefont
  {{Freire}}}, \bibinfo {author} {\bibfnamefont {J.~W.~T.}\ \bibnamefont
  {{Hessels}}}, \bibinfo {author} {\bibfnamefont {I.~H.}\ \bibnamefont
  {{Stairs}}}, \bibinfo {author} {\bibfnamefont {P.}~\bibnamefont {{Arras}}},\
  and\ \bibinfo {author} {\bibfnamefont {M.}~\bibnamefont {{Cadelano}}},\
  }\bibfield  {title} {\bibinfo {title} {{Using Long-term Millisecond Pulsar
  Timing to Obtain Physical Characteristics of the Bulge Globular Cluster
  Terzan 5}},\ }\href {https://doi.org/10.3847/1538-4357/aa7ed7} {\bibfield
  {journal} {\bibinfo  {journal} {\apj}\ }\textbf {\bibinfo {volume} {845}},\
  \bibinfo {eid} {148} (\bibinfo {year} {2017})},\ \Eprint
  {https://arxiv.org/abs/1612.04395} {arXiv:1612.04395 [astro-ph.SR]}
  \BibitemShut {NoStop}%
\bibitem [{\citenamefont {{Blandford}}\ \emph {et~al.}(1987)\citenamefont
  {{Blandford}}, \citenamefont {{Romani}},\ and\ \citenamefont
  {{Applegate}}}]{1987MNRAS.225P..51B}%
  \BibitemOpen
  \bibfield  {author} {\bibinfo {author} {\bibfnamefont {R.~D.}\ \bibnamefont
  {{Blandford}}}, \bibinfo {author} {\bibfnamefont {R.~W.}\ \bibnamefont
  {{Romani}}},\ and\ \bibinfo {author} {\bibfnamefont {J.~H.}\ \bibnamefont
  {{Applegate}}},\ }\bibfield  {title} {\bibinfo {title} {{Timing a millisecond
  pulsar in a globular cluster.}},\ }\href
  {https://doi.org/10.1093/mnras/225.1.51P} {\bibfield  {journal} {\bibinfo
  {journal} {\mnras}\ }\textbf {\bibinfo {volume} {225}},\ \bibinfo {pages}
  {51P} (\bibinfo {year} {1987})}\BibitemShut {NoStop}%
\bibitem [{\citenamefont {{Phinney}}(1992)}]{1992RSPTA.341...39P}%
  \BibitemOpen
  \bibfield  {author} {\bibinfo {author} {\bibfnamefont {E.~S.}\ \bibnamefont
  {{Phinney}}},\ }\bibfield  {title} {\bibinfo {title} {{Pulsars as Probes of
  Newtonian Dynamical Systems}},\ }\href
  {https://doi.org/10.1098/rsta.1992.0084} {\bibfield  {journal} {\bibinfo
  {journal} {Philosophical Transactions of the Royal Society of London Series
  A}\ }\textbf {\bibinfo {volume} {341}},\ \bibinfo {pages} {39} (\bibinfo
  {year} {1992})}\BibitemShut {NoStop}%
\bibitem [{\citenamefont {{Maggiore}}(2007)}]{maggiore07}%
  \BibitemOpen
  \bibfield  {author} {\bibinfo {author} {\bibfnamefont {M.}~\bibnamefont
  {{Maggiore}}},\ }\href
  {https://doi.org/10.1093/acprof:oso/9780198570745.001.0001} {\emph {\bibinfo
  {title} {{Gravitational Waves: Volume 1: Theory and Experiments}}}}\
  (\bibinfo  {publisher} {Oxford University Pres},\ \bibinfo {year}
  {2007})\BibitemShut {NoStop}%
\bibitem [{\citenamefont {{Peters}}\ and\ \citenamefont
  {{Mathews}}(1963)}]{Peters_1963}%
  \BibitemOpen
  \bibfield  {author} {\bibinfo {author} {\bibfnamefont {P.~C.}\ \bibnamefont
  {{Peters}}}\ and\ \bibinfo {author} {\bibfnamefont {J.}~\bibnamefont
  {{Mathews}}},\ }\bibfield  {title} {\bibinfo {title} {{Gravitational
  Radiation from Point Masses in a Keplerian Orbit}},\ }\href
  {https://doi.org/10.1103/PhysRev.131.435} {\bibfield  {journal} {\bibinfo
  {journal} {Physical Review}\ }\textbf {\bibinfo {volume} {131}},\ \bibinfo
  {pages} {435} (\bibinfo {year} {1963})}\BibitemShut {NoStop}%
\bibitem [{\citenamefont {{Ba{\~n}ares-Hern{\'a}ndez}}\ \emph
  {et~al.}(2024)\citenamefont {{Ba{\~n}ares-Hern{\'a}ndez}}, \citenamefont
  {{Calore}}, \citenamefont {{Camalich}},\ and\ \citenamefont
  {{Read}}}]{Banares_2024}%
  \BibitemOpen
  \bibfield  {author} {\bibinfo {author} {\bibfnamefont {A.}~\bibnamefont
  {{Ba{\~n}ares-Hern{\'a}ndez}}}, \bibinfo {author} {\bibfnamefont
  {F.}~\bibnamefont {{Calore}}}, \bibinfo {author} {\bibfnamefont {J.~M.}\
  \bibnamefont {{Camalich}}},\ and\ \bibinfo {author} {\bibfnamefont {J.~I.}\
  \bibnamefont {{Read}}},\ }\bibfield  {title} {\bibinfo {title} {{New
  constraints on the central mass contents of Omega Centauri from combined
  stellar kinematics and pulsar timing}},\ }\href
  {https://doi.org/10.48550/arXiv.2408.00939} {\bibfield  {journal} {\bibinfo
  {journal} {arXiv e-prints}\ ,\ \bibinfo {eid} {arXiv:2408.00939}} (\bibinfo
  {year} {2024})},\ \Eprint {https://arxiv.org/abs/2408.00939}
  {arXiv:2408.00939 [astro-ph.GA]} \BibitemShut {NoStop}%
\bibitem [{\citenamefont {{Gerssen}}\ \emph {et~al.}(2002)\citenamefont
  {{Gerssen}}, \citenamefont {{van der Marel}}, \citenamefont {{Gebhardt}},
  \citenamefont {{Guhathakurta}}, \citenamefont {{Peterson}},\ and\
  \citenamefont {{Pryor}}}]{2002AJ....124.3270G}%
  \BibitemOpen
  \bibfield  {author} {\bibinfo {author} {\bibfnamefont {J.}~\bibnamefont
  {{Gerssen}}}, \bibinfo {author} {\bibfnamefont {R.~P.}\ \bibnamefont {{van
  der Marel}}}, \bibinfo {author} {\bibfnamefont {K.}~\bibnamefont
  {{Gebhardt}}}, \bibinfo {author} {\bibfnamefont {P.}~\bibnamefont
  {{Guhathakurta}}}, \bibinfo {author} {\bibfnamefont {R.~C.}\ \bibnamefont
  {{Peterson}}},\ and\ \bibinfo {author} {\bibfnamefont {C.}~\bibnamefont
  {{Pryor}}},\ }\bibfield  {title} {\bibinfo {title} {{Hubble Space Telescope
  Evidence for an Intermediate-Mass Black Hole in the Globular Cluster M15. II.
  Kinematic Analysis and Dynamical Modeling}},\ }\href
  {https://doi.org/10.1086/344584} {\bibfield  {journal} {\bibinfo  {journal}
  {\aj}\ }\textbf {\bibinfo {volume} {124}},\ \bibinfo {pages} {3270} (\bibinfo
  {year} {2002})},\ \Eprint {https://arxiv.org/abs/astro-ph/0209315}
  {arXiv:astro-ph/0209315 [astro-ph]} \BibitemShut {NoStop}%
\bibitem [{\citenamefont {{Baumgardt}}\ \emph {et~al.}(2003)\citenamefont
  {{Baumgardt}}, \citenamefont {{Hut}}, \citenamefont {{Makino}}, \citenamefont
  {{McMillan}},\ and\ \citenamefont {{Portegies Zwart}}}]{2003ApJ...582L..21B}%
  \BibitemOpen
  \bibfield  {author} {\bibinfo {author} {\bibfnamefont {H.}~\bibnamefont
  {{Baumgardt}}}, \bibinfo {author} {\bibfnamefont {P.}~\bibnamefont {{Hut}}},
  \bibinfo {author} {\bibfnamefont {J.}~\bibnamefont {{Makino}}}, \bibinfo
  {author} {\bibfnamefont {S.}~\bibnamefont {{McMillan}}},\ and\ \bibinfo
  {author} {\bibfnamefont {S.}~\bibnamefont {{Portegies Zwart}}},\ }\bibfield
  {title} {\bibinfo {title} {{On the Central Structure of M15}},\ }\href
  {https://doi.org/10.1086/367537} {\bibfield  {journal} {\bibinfo  {journal}
  {\apjl}\ }\textbf {\bibinfo {volume} {582}},\ \bibinfo {pages} {L21}
  (\bibinfo {year} {2003})},\ \Eprint {https://arxiv.org/abs/astro-ph/0210133}
  {arXiv:astro-ph/0210133 [astro-ph]} \BibitemShut {NoStop}%
\end{thebibliography}%



\end{document}